\pdfoutput=1

\documentclass[conference]{IEEEtran}
\IEEEoverridecommandlockouts
\usepackage{graphicx}
\usepackage{balance}  
\usepackage{xcolor}
\usepackage{amsfonts,amssymb,amsmath,bm}
\usepackage{graphicx}
\usepackage{makecell}
\usepackage{subfigure}
\usepackage{multirow}
\usepackage[ruled,vlined]{algorithm2e}
\usepackage{verbatim}

\newtheorem{example}{Example}
\newtheorem{definition}{Definition}

\begin{document}


\title{KGClean: An Embedding Powered Knowledge Graph Cleaning Framework}

\author{
    \IEEEauthorblockN{Congcong Ge$^{\dagger}$, Yunjun Gao$^{\dagger}$$^{\sharp}$, Honghui Weng$^{\dagger}$, Chong Zhang$^{\dagger}$, Xiaoye Miao$^{\ddagger}$, Baihua Zheng$^{*}$}
    \IEEEauthorblockA{
        \textit{{\large$^{\dagger}$}College of Computer Science, Zhejiang University, Hangzhou, China} \\
        \textit{{\large$^{\ddagger}$}Center for Data Science, Zhejiang University, Hangzhou, China} \\
        \textit{{\large$^{\sharp}$}Alibaba--Zhejiang University Joint Institute of Frontier Technologies, Hangzhou, China} \\
        \textit{{\large$^{*}$}School of Information Systems, Singapore Management University, Singapore, Singapore} \\
        {\large$^{\dagger}$}{\large$^{\ddagger}$}\{gcc, gaoyj, wenghh, zhangchong, miaoxy\}@zju.edu.cn~~~~{\large$^{*}$}bhzheng@smu.edu.sg
    }
}

\maketitle



\begin{abstract}
The quality assurance of the knowledge graph is a prerequisite for various knowledge-driven applications. We propose \textsf{KGClean}, a novel cleaning framework powered by knowledge graph embedding, to detect and repair the heterogeneous dirty data. In contrast to previous approaches that either focus on filling missing data or clean errors violated limited rules, \textsf{KGClean} enables (i) cleaning both missing data and other erroneous values, and (ii) mining potential rules automatically, which expands the coverage of error detecting. \textsf{KGClean} first learns \emph{data representations} by \textsf{TransGAT}, an effective knowledge graph embedding model, which gathers the neighborhood information of each data and incorporates the interactions among data for casting data to continuous vector spaces with rich semantics.
\textsf{KGClean} integrates an active learning-based classification model, which \emph{identifies} errors with a small seed of labels. KGClean utilizes an efficient PRO-repair strategy to \emph{repair} errors using a novel concept of \emph{propagation power}. Extensive experiments on four typical knowledge graphs demonstrate the effectiveness of \textsf{KGClean} in practice.
\end{abstract}

\section{Introduction}
\label{sec:intro}

\begin{figure}[t]
\centering
\includegraphics[width=3.1in]{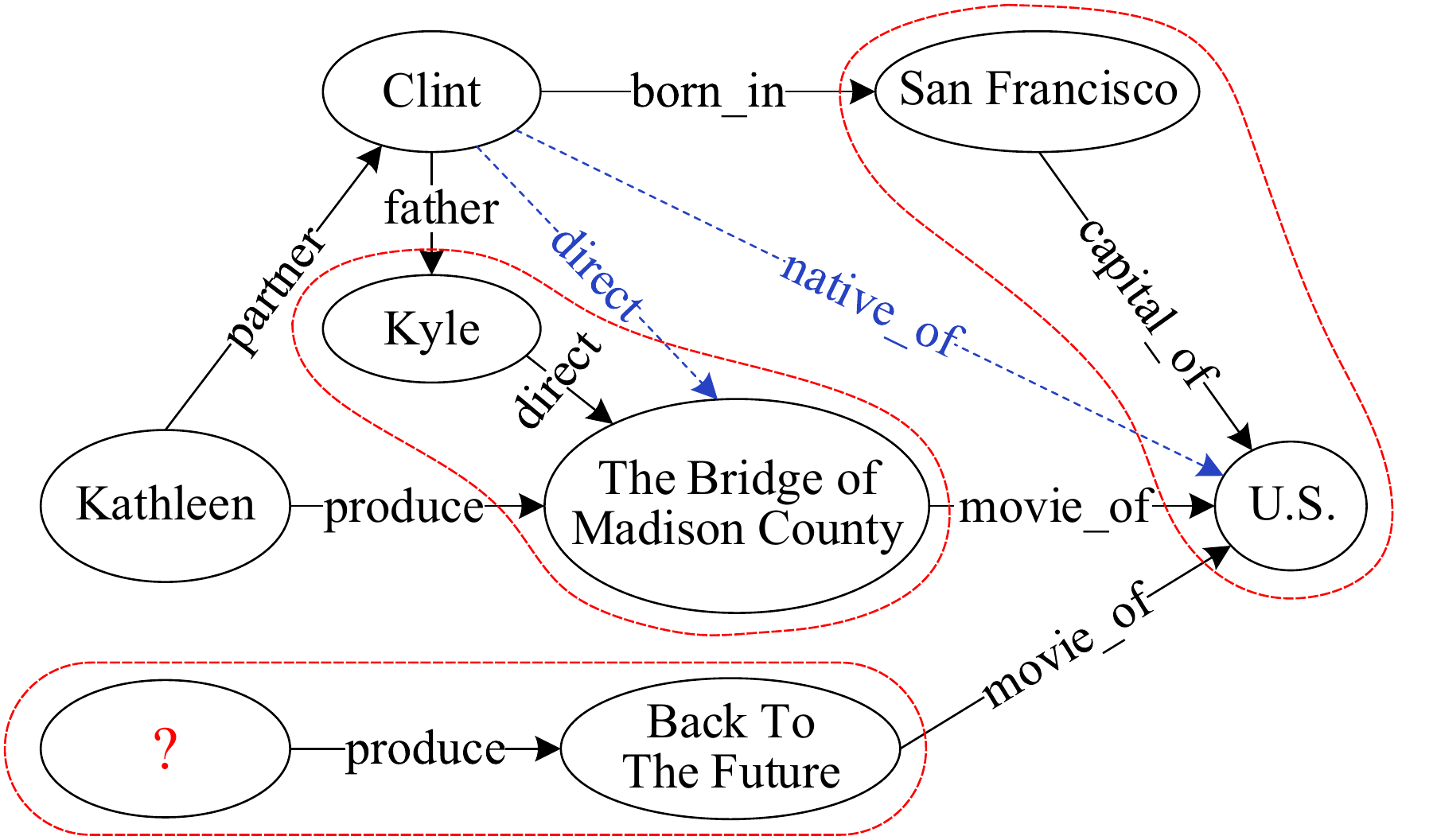}
\vspace*{-2mm}
\caption{A sample of a KG with dirty data}
\label{fig:intro}
\vspace*{-4mm}
\end{figure}

Knowledge graph (KG) is a widely used human-knowledge representation model, which consists of \emph{entities} and their rich \emph{relationships}.
A KG can be treated as a set of triplets.
For example, a triplet (Clint, born\_in, San Francisco) in Figure~\ref{fig:intro} is represented as two entities: ``Clint'' and ``San Francisco'', along with a relationship ``born\_in'' linking them.
Each node denotes an entity, and each arrow with a solid line indicates a relationship between two entities.
KGs motivate many knowledge-driven scenarios, such as semantic search~\cite{BerantL14, GivoliR19} and question answering~\cite{DiefenbachSM18, ZhongXKS19}, to name but a few.
However, KGs suffer from several knowledge quality problems. As an example, most of KGs are automatically extracted from web sources, and the precision might be low, e.g., the estimated precision of NELL, a never-ending language learning system that learns over time to read the web,
is only 74\%~\cite{CarlsonBKSHM10}.
Although knowledge graph completion methods~\cite{MeilickeCRS19, SocherCMN13, XueYXS18} have been proposed to improve the quality of KGs, they only aim at the task of predicting missing entities but not the task of recognizing or correcting the erroneous values, including both missing values and wrong values. We would like to highlight that the capability of denoising KGs is even more important than completing KGs, because a KG containing erroneous values is more likely to generate wrong answers.

\vspace*{-1mm}
\begin{example}
Figure \ref{fig:intro} depicts a sample knowledge graph (KG) containing dirty data (highlighted with red circles), including wrong values and missing values.
For instance, (?, produce, Back To The Future) is a triplet that includes a missing value in the head, and knowledge graph completion methods can fix it. Triplets (Kyle, direct, The Bridge of Madison County) and (San Francisco, capital\_of, U.S.) contain wrong values, since ``Kyle'' is not the director of the movie called ``The Bridge of Madison County'', and ``San Francisco'' is not the capital of ``U.S.''. Take the triplet (San Francisco, capital\_of, U.S.) as an example, we expect it to be repaired by either modifying the relationship value ``capital\_of'' to ``city\_of'' or replacing the entity value ``San Francisco'' with ``Washington, D.C.''. Unfortunately, existing knowledge graph completion methods cannot identify the wrong values of entities or relationships from these triples, and let alone tell us which modification is more suitable for this sample.
\end{example}

An important and effective way to further improve the data quality of KGs is to clean dirty data, i.e., \emph{detecting} and \emph{repairing} erroneous values. The former refers to the capability of identifying erroneous values, and the latter refers to the capability of correcting the erroneous values.

In the field of general graphs, several error detecting \cite{CalvaneseFPSS14,CalabuigP12,FanL17,FanWX16a,YuH11} or repairing approaches \cite{FanLTZ19} have been developed to identify or fix errors for graph data.
Those methods detect or repair errors that violate data quality rules. Nonetheless, getting sufficient data quality rules is labor-intensive, and thus makes those cleaning methods less effective.
When a graph data is intricate, it is unrealistic to assume that all the required rules could be identified.
Therefore, it is hard to repair errors that are not included in any of the identified rules.

One has witnessed an increasing availability of \emph{knowledge graph embedding} techniques, whose purpose is to transform words of the corpus into different values in vector spaces retaining their semantic information.
Knowledge graph embedding requires that every triplet of a KG should obey the \emph{causality} (can be treated as a weighted rule). If a triplet holds, it has a strong causality. On the contrary, if the causality of a triplet is weak, the triplet is likely to be erroneous. Furthermore, knowledge graph embedding is able to not only automatically measure the strength of the causalities followed by the given triplets, but also learn the causalities that potential triplets should obey.

\begin{example}
The arrows with solid lines in Figure \ref{fig:intro} denote relationships between entities.
An arrow with a dashed line represents a possible relationship that can be inferred from the known entities.
For instance, the embedding model discovers three strong causalities from the given KG, i.e., (i) Kathleen is the producer of the movie ``The Bridge of Madison County'', (ii) Kathleen is Clint's partner, and (iii) Clint is Kyle's father.
In view of the semantic information contained in the above triplets, the embedding model may infer that, Clint is likely to be the director of the movie ``The Bridge of Madison County''. This newly inferred causality can help identify the erroneous triplet (Kyle, direct, The Bridge of Madison County), and replace entity ``Kyle'' with entity ``Clint'' in the triplet as a repairing.
\end{example}

Motivated by the power of knowledge graph embedding and the potential close relationship between erroneous data in KGs and weak causality in the embedding space, we propose to perform the task of cleaning dirty data with the help of knowledge graph embedding. To be more specific, after a given KG is transformed into a vector space, the task of \textit{error detecting} could be redefined as a binary classification problem, i.e., whether a triplet shall be classified into a dirty class or a clean class; and the task of \textit{error repairing} could be redefined as a probability inference problem, i.e., repairing erroneous values in dirty triplets to make the probability of each repaired triplet as high as possible.
%
%

\noindent
\textbf{Challenges.}
%
The success of the above-mentioned idea largely depends on whether we are able to effectively address following challenges.
\vspace*{-1mm}
\begin{itemize}\setlength{\itemsep}{-\itemsep}
  \item \textbf{[Embedding.]} Heterogeneity of dirty data makes it challenging to detect and repair dirty data accurately on KGs. Hence, we need a novel knowledge graph embedding model that can appropriately express knowledge graph data into semantic spaces.
  \item \textbf{[Annotation.]} Getting a reliable classification model for a KG requires sufficient correctly-labeled training data. Since the annotation process is labor-intensive, it is desirable to reduce the demand on labeled data while ensuring the reliability of the model.
  \item \textbf{[Interpretability.]} The combination of data cleaning and embedding requires interpretability. Embedding is often viewed as a black box that is hard to explain. Nevertheless, data cleaning needs explainable reasons to guide the analyses of how errors may occur or be repaired.
\end{itemize}

To address these challenges, we present \textsf{KGClean}, a data cleaning framework that leverages a novel embedding model, for detecting and repairing both dirty entities and dirty relationships on KGs. \textsf{KGClean} first introduces \textsf{TransGAT}, an effective KG embedding model that leverages a high-quality external corpus, to learn representations of entities and relationships in vector spaces.
Then, \textsf{KGClean} utilizes a newly presented \textsf{AL-detect} strategy to classify triplets as either clean or dirty, by incorporating active learning techniques.
Finally, \textsf{KGClean} adopts an error repairing strategy using a novel concept of propagation power (\textsf{PRO-repair} for short) to clean erroneous values within dirty triplets.

\noindent
\textbf{Contributions.} We summarize the key contributions of \textsf{KGClean} as follows:
\vspace*{-1mm}
\begin{itemize}\setlength{\itemsep}{-\itemsep}
    \item \textit{A knowledge graph cleaning framework.} We propose the first knowledge graph cleaning framework that is powered by knowledge graph embedding, which aims to detect and clean the heterogeneous dirty data (including both entities and relationships) within a KG.
    \item \textit{Rich semantic information for embedding.} We introduce a novel knowledge graph embedding model, i.e., \textsf{TransGAT}, to cast entities and relationships to vector spaces. \textsf{TransGAT} obtains rich semantic information for entities and relationships by (i) gathering information from the entities' neighbors as well as (ii) considering the interactions between entities and relationships. The appropriate embedding model of \textsf{TransGAT} guarantees the accuracy of the subsequent error detecting and error repairing stages of knowledge graph cleaning.
    \item \textit{Fewer human involvement in data annotation.} We design \textsf{AL-detect}, an active learning-based classification method to identify errors, which reduces the number of labels by filtering unimportant triplets.
    \item \textit{Interpretable reasons for cleaning.} We present a \textsf{PRO-repair} method for cleaning erroneous values in noisy triplets and meanwhile making the repaired values interpretable. The underlying idea is that, if errors are detected based on an explicit path on a KG, it will be easy to interpret the cause of the cleaning result with the support of causal evidence on the path.
\end{itemize}

\noindent
\textbf{Organization.}
The rest of this paper is organized as follows. Section \ref{sec:related_work} reviews the related work. Section \ref{sec:back_ground} covers the basic background materials and techniques used in the paper. Section \ref{sec:framework_overview} overviews our cleaning framework \textsf{KGClean}. Section \ref{sec:representation} describes the representations of entities and relationships on the knowledge graph. Section \ref{sec:data_cleaning} elaborates error detecting and error repairing in the data cleaning process. Section \ref{sec:experiments} reports the experimental results and our findings. Finally, Section \ref{sec:conclusion} concludes the paper.

\section{related work}
\label{sec:related_work}
\noindent
\textbf{Data Cleaning.}
There has been a surge of interest in data cleaning from industry and academia \cite{CleaningSurvey16}.
Existing data cleaning methods and prototypes can be classified into four categories, i.e., (i) KG powered cleaning methods, (ii) rule-based cleaning methods, (iii) statistical cleaning methods, and (iv) user (experts or crowd) interaction cleaning ones.

In the first category, KGs are used as evidence for identifying errors that mismatch values in KGs.
KATARA \cite{KATARA15} uses crowdsourcing as a complement to verify whether values that mismatch KGs are correct or not.
\cite{DRs18} introduces new declarative rules (DRs) to model the relationship between KGs and clean the data without the involvement of humans. One main difference between \textsf{KGClean} and existing KG powered cleaning systems is that, \textsf{KGClean} cleans KGs directly to improve their data quality, while others consider KGs as \emph{clean} external information to support the data cleaning process for structured datasets.

Rule-based methods are classic in data cleaning. For structured datasets, they clean errors that violate integrity constraint rules, such as ones using functional dependencies (FDs) \cite{sampling10, BeskalesIGG13, BohannonFFR05, BigDansing15, KolahiL09, UniDetect19}, conditional functional dependencies (CFDs) \cite{BohannonFGJK07, FanGJK08, LLUNATIC13, BigDansing15}, and denial constraints (DCs) \cite{Holistic13, LLUNATIC13, HoloDetect, BigDansing15, LopatenkoB07, HoloClean17}, to name just a few. 
For graph datasets, many attempts are made to detect errors \cite{CalvaneseFPSS14, CalabuigP12, FanL17, FanWX16a, YuH11} or repair errors \cite{FanLTZ19} based on graph quality rules. Given the correct rules, those studies clean data that violates the rules, but they are not able to detect errors that are not contained by the rules. However, they are all limited by the difficulty of obtaining sufficient and correct rules. 
On the other hand, they are orthogonal to \textsf{KGClean} because of the elimination of dependency on the given graph quality rules. \textsf{KGClean} employs the knowledge graph embedding model to automatically learn causalities, which could be considered as rules that can guide value cleaning in a knowledge graph. 

Statistical cleaning methods  repair errors based on data probabilistic distributions \cite{HoloDetect, ActiveClean16, Raha19, CurrentClean19, HoloClean17, UniDetect19}. Existing statistical techniques are designed for structured data cleaning, whereas \textsf{KGClean} focuses on using \textsf{TransGAT}, a deep learning-based KG embedding model, for knowledge graph cleaning.

User interaction cleaning methods utilize human knowledge to improve the quality of cleaning results on the premise of budget minimization \cite{DANCE17, KATARA15, HeVSLMPT16, ActiveClean16, Raha19, UGuide17, CrowdCleaner14, YakoutENOI11}.
They could be considered as a complement to, but not a competitor of, \textsf{KGClean}.


\noindent
\textbf{Distributed Representations.}
Distributed representations are sets of dense vectors with low dimensions that can describe the semantic similarities among textual values. The mainstream distributed representations are categorized into word embeddings and  graph embeddings.

Word embeddings assign an appropriate vector to each textual value, based on its context information of the sentence to which the value belongs, e.g., word2vec~\cite{word2vec13a, word2vec13b}, GloVe~\cite{glove2014}, fastText~\cite{fastText17}, etc.
Graph embeddings learn vector representations based on the relationships among nodes on graphs, including using the techniques of random walk \cite{node2vec16, DeepWalk14}, factorization \cite{AhmedSNJS13, BelkinN01, OuCPZ016, roweis2000nonlinear}, and deep-learning \cite{GCN17, GAT18, SDNE16}. We focus on knowledge graph embedding \cite{TransE13, TransR15, NathaniCSK19}, a kind of graph embedding models, to learn the embeddings of entities and relationships of the knowledge graph.
Closer to our work is \cite{NathaniCSK19}, one of the state-of-the-art KG embedding models. It captures the features of entities and relationships in each specified entity's neighborhood to make the learned entity embeddings contain rich semantic information. Different from \cite{NathaniCSK19}, our proposed KG embedding model \textsf{TransGAT} not only uses the relationship features to enrich the semantic information of entity embeddings, but also applies the information contained by entities to learn relationship embeddings.


\section{PRELIMINARIES}
\label{sec:back_ground}
In this section, we describe some background materials and techniques used in sections later. Table \ref{table:notation} summarizes the symbols used frequently throughout this paper.

\begin{table}[t]
\vspace*{-2mm}
\caption{Symbols and Description}
\label{table:notation}
\setlength{\tabcolsep}{5pt}
\begin{tabular}{|c|p{6cm}|}
\hline
\textbf{Notation} & \textbf{Description}                                                \\ \hline
$\mathcal{G}$   & a knowledge graph                                                     \\
$\mathcal{E}$   & a set of entities in $\mathcal{G}$                                    \\
$\mathcal{R}$   & a set of relationship types in $\mathcal{G}$                                   \\
$e_h$           & a head entity value belonging to $\mathcal{E}$                          \\
$e_t$           & a tail entity value belonging to $\mathcal{E}$                          \\
$r_k$           & a relationship value belonging to $\mathcal{R}$                             \\
$t(e_h,r_k,e_t)$   & a triplet denoting an edge $r_{k}$ from $e_h$ to $e_t$                  \\
$\bm{e_h} \in \mathbb{R}^{N_h \times F_e}$   & the embeddings for head entities in $\mathcal{E}$   \\
$\bm{e_t} \in \mathbb{R}^{N_t \times F_e}$   & the embeddings for tail entities in $\mathcal{E}$   \\
$\bm{r_k} \in \mathbb{R}^{N_r \times F_r}$   & the embeddings for relationships in $\mathcal{R}$       \\
\hline
\end{tabular}
\vspace*{-4mm}
\end{table}

\begin{figure*}[t]
\centering
\includegraphics[width=7.15in]{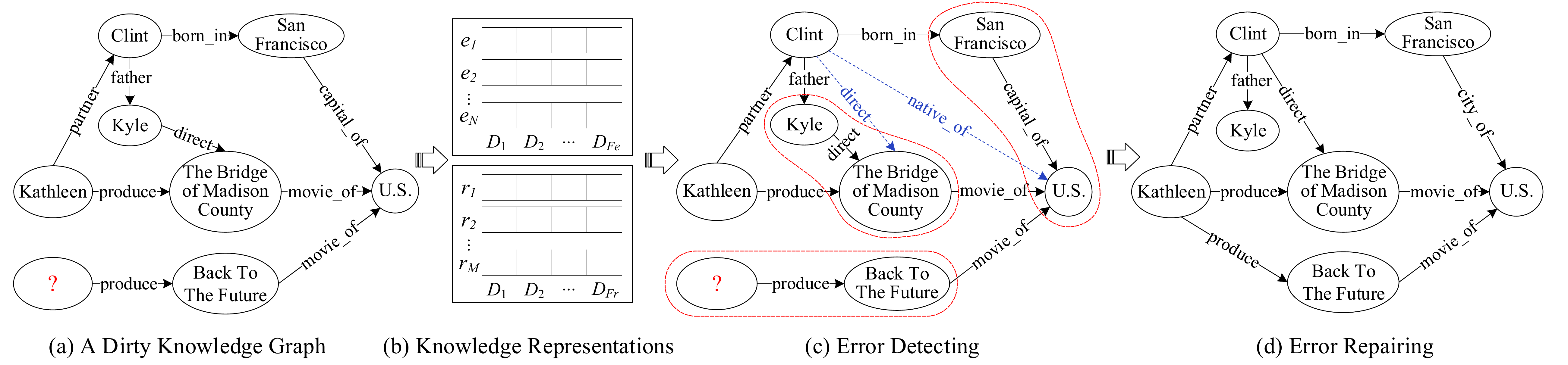}
\vspace*{-5.5mm}
\caption{Workflow diagram of KGClean Framework}
\label{fig:framework}
\vspace*{-2mm}
\end{figure*}

\subsection{Knowledge Graph Embedding}
A knowledge graph (KG) is a multi-relational directed graph, denoted as $\mathcal{G}=(\mathcal{E}, \mathcal{R})$, where $\mathcal{E}$ and $\mathcal{R}$ represent the set of entities (w.r.t. nodes) and set of relationship types (w.r.t. edges), respectively. A triplet $t = (e_h, r_{k}, e_t)$ represents an edge $r_{k}$ from a head node $e_h$ to a tail node $e_t$ in $\mathcal{G}$. Without loss of generality, we use the terms of ``node/entity'' and ``edge/relationship'' interchangeably throughout the paper. Different from general graphs, edges associated with two triplets $t_1$ and $t_2$ could be the same, i.e., $t_1.r_{k}=t_2.r_{k}$, as we differentiate an edge from another based on the type but not their head entity or tail entity.

The entities and relationships in KG are widely stored as textual values, and the underlying symbolic nature of such values usually makes KGs hard to manipulate. A simple way is to encode entities and relationships with one-hot vectors, but it may cause dimension explosion and the lack of semantics. The key idea of knowledge graph (KG) embedding is to map entities and relationships of a KG into compressed continuous vector spaces retaining their semantic information. A general KG embedding follows three steps: (i) representing entities $\mathcal{E}$ and relationships $\mathcal{R}$, (ii) defining a score function $f_r$, and (iii) learning an effective representation of $\mathcal{E}$ and $\mathcal{R}$ based on the score function~\cite{KGSurvey17}.

\subsection{Graph Attention Networks}

Graph attention networks (GATs)~\cite{GAT18} are novel neural network architectures that operate on graph-structured data.
GATs learn to assign appropriate vectors to entities by taking into account the information from their neighbors.

A graph attentional layer takes a set of nodes, $\bm{e}=$\{$\vec{e_1}$, $\vec{e_2}$, ..., $\vec{e_N}$\}$ \in \mathbb{R}^{N \times F_e}$ as an input, where $N$ is the number of nodes, and $F_e$ is the feature dimensionality of each node embedding.
The layer outputs a new set of nodes, $\bm{e'}=$\{$\vec{e'_1}$, $\vec{e'_2}$, ..., $\vec{e'_N}$\} $\in \mathbb{R}^{N \times F_{e'}}$, in which $F_{e'}$ is the new feature dimensionality of each node embedding. A single GAT layer is given by
\begin{equation}
\gamma_{ij}=a(\mathbf{W}\vec{e_i}, \mathbf{W}\vec{e_j})
\end{equation}
where $\gamma_{ij}$ is the \emph{attention coefficient} that indicates the importance of node $e_i$ to node $e_j$ with $e_j$ being a neighbor of $e_i$, $\mathbf{W}$ is a parameterized linear transformation matrix, and $a$ is a chosen attentional function.

To make attention coefficients easily comparable across different nodes, the relative attention coefficient is computed using a \emph{softmax function}, as shown in Equation~(\ref{eq:relative-attention}).

\begin{equation}
\label{eq:relative-attention}
\alpha_{ij}={\rm softmax}_j (\gamma_{ij})=\frac{{\rm exp}(\gamma_{ij})}{\sum_{e_{k} \in \mathcal{N}_i} {\rm exp}(\gamma_{ik})}
\end{equation}
where $\mathcal{N}_i$ denotes the set of neighbors of $e_i$.
After $\alpha_{ij}$ is derived, the output embedding can be calculated with a nonlinearity $\sigma$, as depicted in Equation~(\ref{eq:gat_output}).
\begin{equation}
\label{eq:gat_output}
\vec{e'_i}=\sigma \left(\sum_{e_j \in \mathcal{N}_i}\alpha_{ij} \mathbf{W} \vec{e_j}\right)
\end{equation}

\subsection{Multi-hop Neighbors}
\label{sec:multi_hop}

Multi-hop neighbors~\cite{NathaniCSK19}, as formally defined in Definition~\ref{defn:mhn}, are proposed to enrich the neighbors' information in learning GATs.
\begin{definition}\label{defn:mhn}
\textbf{(Multi-hop Neighbors).}
Given a triplet $(e_{L-1},$ $r_{L},$ $e_{L})$, an edge $r_{L}$ from node $e_{L-1}$ to node $e_{L}$ can be represented as a path $e_{L-1} \stackrel{r_{L}}{\longrightarrow} e_{L}$, where $e_{L-1}$ is a direct in-flowing neighbor of $e_{L}$, also defined as a 1-hop neighbor of $e_{L}$.
More generally, given a path $e_{0} \stackrel{r_{1}}{\longrightarrow} e_{1} \stackrel{r_{2}}{\longrightarrow} \cdots \stackrel{r_{L}}{\longrightarrow} e_{L}$, $e_{0}$ is defined as an $L$-hop ($L>1$) neighbor of $e_{L}$.
\end{definition}

All entities in a KG $\mathcal{G}$ capture information from their multi-hop neighbors.
Given two entities $e_{0}$ and $e_{L}$, where $e_{0}$ is an $L$-hop ($L>1$) neighbor of $e_{L}$, we denote an auxiliary edge that directly connects $e_{0}$ to $e_{L}$ as a triplet $(e_{0}, r_{aux}, e_{L})$, where $r_{aux}$ is a potential relationship between $e_{0}$ and $e_{L}$. Take Figure~\ref{fig:intro} as an example, entity ``Clint'' is a 2-hop neighbor of entity ``The Bridge of Madison County'', and the potential relationship between them is ``direct''. Also, entity ``Clint'' is a 2-hop neighbor of entity ``U.S.'', and the potential relationship between them is ``native\_of''.
We use $\mathcal{R}_{aux}$ to represent the set of potential relationships.
Thus, the KG $\mathcal{G}$ is enriched from $\mathcal{G}=\{\mathcal{E}, \mathcal{R}\}$ to $\mathcal{G}'=\{\mathcal{E}, \mathcal{R}'\}$, where $\mathcal{R}' = \mathcal{R} + \mathcal{R}_{aux}$.

\section{Framework Overview}
\label{sec:framework_overview}

In this section, we first formalize the problem of data cleaning in Section~\ref{sec:problem_statement}, and then overview the framework of \textsf{KGClean} in Section~\ref{subsec:framework_overview}.

\subsection{Problem Statement}
\label{sec:problem_statement}

We assume that errors in a dirty knowledge graph $\mathcal{G}_d$ occur due to inaccurate value assignments, a common assumption made by many data cleaning systems~\cite{Holistic13, KATARA15, HoloDetect, HoloClean17}. The goal of \textsf{KGClean} is to detect and repair both erroneous entities and erroneous relationships, including missing values and wrong values, in $\mathcal{G}_d$, according to the embeddings of entities and relationships learned from a training knowledge graph $\mathcal{G}_t$ that contains only clean entities and relationships. 
$\mathcal{E}=\{e_1,e_2,...,e_N\}$ denotes the entities of a knowledge graph, where $N$ is the number of entities. 
$\mathcal{R}=\{r_1,r_2,...,r_M\}$ represents the relationship types of a knowledge graph, where $M$ is the number of relationship types. Each relationship $r_k$ from an entity $e_h$ to another entity $e_t$ is represented as a triplet, denoted by $(e_h, r_k, e_t)$.

\subsection{KGClean Overview}
\label{subsec:framework_overview}

The workflow of \textsf{KGClean} is illustrated in Figure \ref{fig:framework}. It is composed of three modules, i.e., (i) knowledge representations; (ii) error detecting; and (iii) error repairing, as detailed below. 
\\
\textbf{Knowledge representations.}
Given a dirty KG $\mathcal{G}_d$ and a training KG $\mathcal{G}_t$ that contains only clean and accurate triplets, the first step of \textsf{KGClean} is to learn the embeddings of entities and relationships in $\mathcal{G}_d$ according to $\mathcal{G}_t$. 
Reliable embeddings are the basis for the subsequent data cleaning process. In order to achieve this, we introduce \textsf{TransGAT} for learning vectors of entities and relationships. (i) \textsf{TransGAT} integrates GATs to enrich the semantic information of entity embeddings according to the neighborhood information. 
(ii) \textsf{TransGAT} considers the \emph{interactions} between entities and relationships when learning their embeddings. The interaction means that the semantic information of entities can facilitate the learning of relationships' embedding, and in turn, the semantic information of relationships can also enhance the embedding of entities. (iii) \textsf{TransGAT} obeys the causality within each triplet to ensure the accuracy of the training results. (iv) \textsf{TransGAT} employs ConvKB \cite{ConvKB2018} to optimize the global embedding properties of each triplet. 
Figure~\ref{fig:framework}(b) depicts an example of the learned embeddings, where entities and relationships are expressed as $|F_e|$ dimensional and $|F_r|$ dimensional vectors, respectively.

\noindent
\textbf{Error detecting.}
\textsf{KGClean} handles error detecting as a binary classification problem.
Given a dirty knowledge graph $\mathcal{G}_d$, whose entities and relationships are represented as embeddings in vector spaces, it detects errors by clustering the triplets in $\mathcal{G}_d$ into \emph{noisy} and \emph{clean} categories, denoted as $\mathcal{G}_d^n$ and $\mathcal{G}_d^c = \mathcal{G}_d \setminus \mathcal{G}_d^n$, respectively.
To perform an accurate classification, we present \textsf{AL-detect}, an active learning-based classification model to classify triplets as either dirty or clean. It iteratively selects and annotates the most informative unlabeled triplets from a training KG, to reduce the number of annotations and to learn a reliable classification model. 
The classification model takes the dirty KG $\mathcal{G}_d$ as input, and outputs a set of noisy triplets $\mathcal{G}_d^n$ and a set of clean triplets $\mathcal{G}_d^c$. Figure~\ref{fig:framework}(c) shows examples of the noisy triplets bounded by red circles. 


\noindent
\textbf{Error repairing.}
\textsf{KGClean} tackles the error repairing task as a probability inference problem, and replaces the erroneous values of entities and relationships in $\mathcal{G}_d^n$ with their candidates having the highest probability. It proposes a novel concept, namely, \emph{propagation power}, to quantify the probability distribution $P_r(\hat{t})$ for any given candidate $\hat{t}$ of a to-be-repaired noisy triplet $t \in \mathcal{G}_d^n$. With the guidance of propagation power, we propose \textsf{PRO-repair}, an error repair strategy to fix dirty values in noisy triplets. 
%
For each noisy triplet $t \in \mathcal{G}_d^n$, \textsf{PRO-repair} generates a set of candidates $\mathbb{T}$ and then picks the one with the maximum propagation power as the optimal choice for repairing the noisy triplet. Figure~\ref{fig:framework}(d) depicts the cleaning results.

\section{knowledge representations}
\label{sec:representation}

\begin{figure}[t]
\centering
\includegraphics[width=3.39in]{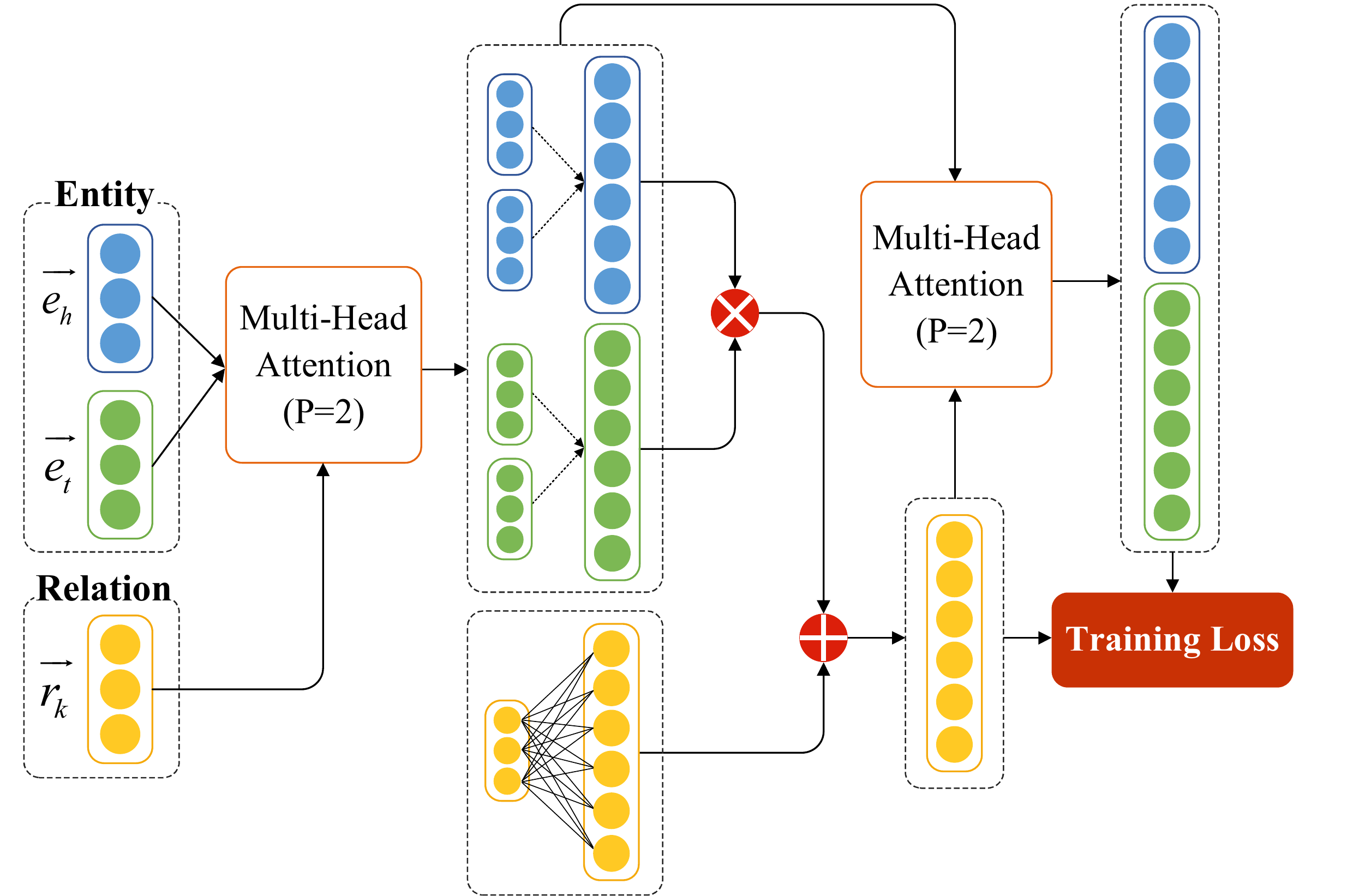}
\vspace*{-5mm}
\caption{TransGAT architecture}
\label{fig:transGAT_model}
\vspace*{-4mm}
\end{figure}

In this section, we present the knowledge graph embedding model \textsf{TransGAT} with two layers to represent entities and relationships using \emph{rich semantic} and \emph{accurate} vectors. The architecture of \textsf{TransGAT} is shown in Figure~\ref{fig:transGAT_model}.

\subsection{The Proposed TransGAT}

To obtain new embeddings for entities and relationships containing \emph{rich semantic information}, \textsf{TransGAT} (i) gathers information from the entities' neighbors using GATs \cite{GAT18}, and (ii) considers the interactions between entities and relationships throughout the two-layer model.

Given a clean knowledge graph $\mathcal{G}_t=\{\mathcal{E}_t, \mathcal{R}_t\}$, we take two randomly initialized embedding matrices as input, i.e., entity embeddings and relationship embeddings. Entity embeddings are represented as a matrix $\bm{e} \in \mathbb{R}^{N_e \times F_e}$, where $N_e$ is the total number of entities and $F_e$ is the feature dimensionality of entity embeddings. Since $\bm{e}$ consists of head embeddings and tail embeddings, we denote head embeddings as a matrix $\bm{e_h} \in \mathbb{R}^{N_h \times F_e}$ and tail embeddings as a matrix $\bm{e_t} \in \mathbb{R}^{N_t \times F_e}$, where $N_h$ and $N_t$ are the number of head entities and that of tail entities respectively, with $N_e=N_h+N_t$.
Relationship embeddings is a matrix $\bm{r} \in \mathbb{R}^{N_r \times F_r}$, where $N_r$ is the number of relationships, and $F_r$ is the feature dimensionality of each relationship embedding.

In the first layer, we input the initialized embedding matrices of entities and relationships and then use neighbors’ information to learn the entity embeddings by applying the attention mechanism of GATs. In the second layer, we first receive the learned entity embeddings from the previous layer, and use the semantics of them for updating the relationship embeddings, and then, we use the updated relationship embeddings to further learn the entity embeddings by applying attention mechanism of GATs again.
After executing the two-layer model, \textsf{TransGAT} then outputs the corresponding embedding matrices, $\bm{e'} \in \mathbb{R}^{N_e \times F_{e'}}$ and $\bm{r'} \in \mathbb{R}^{N_r \times F_{r'}}$.

\noindent
\textbf{Information of neighbors.}
First, we introduce how \textsf{TransGAT} uses neighbors' information to enrich the semantics of entity embeddings.
We extend the set of relationships in $\mathcal{G}_t$ from $\mathcal{R}_t$ to $\mathcal{R}'_t$ using multi-hop neighbors, as stated in Section~\ref{sec:multi_hop}.
Given a triplet $t = (e_h, r_k, e_t)$, where $e_h, e_t \in \mathcal{E}_t$ and $r_k \in \mathcal{R}'_t$, we learn the embeddings of its entities and relationship by performing a linear transformation, as  shown in Equation~(\ref{eq:linearTrans}).
\begin{equation}
\label{eq:linearTrans}
\vec{t}= \mathbf{W_e} (\vec{e_h} + \vec{e_t}) \otimes  \mathbf{W_r} \vec{r_k}
\end{equation}
Here, $\vec{t}$ is the embedding of $t$.
$\mathbf{W_e}$ and $\mathbf{W_r}$ denote the linear transformation matrices corresponding to the entity and relationship, respectively.
Similar to \cite{NathaniCSK19}, we implement $\lambda$ via relational attention mechanism to learn the importance of each triplet $t$, which is formulated as Equation~(\ref{eq:absolute_attention}).
\begin{equation}\label{eq:absolute_attention}
\lambda = \operatorname{LeakyReLU}(\mathbf{W_t} \vec{t})
\end{equation}
where $\mathbf{W_t}$ represents a linear transformation matrix corresponding to $\vec{t}$, and \emph{LeakyReLU} is an activation function.

We normalize the attention value by applying a \emph{softmax} function, as shown in Equation~(\ref{eq:relative_attention}).
\begin{equation}\label{eq:relative_attention}
\alpha= \frac{{\rm exp}(\lambda)}{\sum_{t_j \in \mathcal{A}_h} {\rm exp}(\lambda_j)}
\end{equation}

\noindent
Here, $\mathcal{A}_h$ denotes the adjacent triplets connected to either the head entity $e_h$ or the tail entity $e_t$ of the triplet $t$. Take head entity $e_h$ as an example. For a given $L$-hop ($L \geqslant 1$) neighbor $e_L$ of $e_h$, it is connected to $e_h$ via an edge $e_{L} \stackrel{r'}{\longrightarrow} e_h$, denoted as a triplet $t' = (e_L, r', e_h)$, where $r' \in \mathcal{R}'$.
The triplet $t'\in \mathcal{A}_h$.

Thereafter, a new embedding of the entity $e_h$ is calculated by gathering the normalized attention values across its adjacent triplets, as shown in Equation~(\ref{eq:entity_embedding1}).
\begin{equation}\label{eq:entity_embedding1}
\vec{e'_{h}} = \sigma \left( \sum_{t_j \in \mathcal{A}_h} \alpha_j \vec{t_j} \right)
\end{equation}

With the purpose of stabilizing the learning process and encapsulating more information from the neighborhood, we further integrate the \emph{multi-head attention}~\cite{GAT18} process for training entity embeddings based on Equation~(\ref{eq:entity_embedding1}). Two methods are developed to implement the multi-head attention process, including a \emph{concatenation-based} method and an \emph{average-based} method \cite{GAT18}.

In the first layer of \textsf{TransGAT}, we utilize the concatenation-based multi-head attention process to get the entity embeddings containing rich semantic information from neighbors, as shown in Equation~(\ref{eq:multihead1}).
\begin{equation}\label{eq:multihead1}
\vec{e_{h}'} = \mathop{\parallel}_{p=1}^{P} \sigma \left( \sum_{t_{j}^{p} \in \mathcal{A}_h} \alpha_{j}^{p} \vec{t_{j}^{p}} \right)
\end{equation}
where $\|$ denotes concatenation, $P$ is the total times of independent embedding calculations, and $\alpha_{j}^{p}$ is the normalized attention value computed by the $p$-th attention mechanism.
$\vec{t_{j}^{p}} = \mathbf{W_{e}^{p}} (\vec{e_h} + \vec{e_t}) \otimes  \mathbf{W_{r}^{p}} \vec{r_k}$, where $\mathbf{W_{e}^{p}}$ and $\mathbf{W_{r}^{p}}$ are the corresponding $p$-th linear transformations, respectively. 
In the implementation, \textsf{TransGAT} performs the attention process with two heads (i.e., $P=2$).

In the second layer of \textsf{TransGAT}, we employ the average-based multi-head attention process instead, to get the final entity embeddings, as shown in Equation~(\ref{eq:multihead2}), because the concatenation is no longer sensitive in this layer~\cite{GAT18}.
\begin{equation}\label{eq:multihead2}
\vec{e_{h}'} = \sigma \left( \frac{1}{P} \sum_{p=1}^{P} \sum_{t_{j}^{p} \in \mathcal{A}_h} \alpha_{j}^{p} \vec{t_{j}^{p}} \right)
\end{equation}

\noindent
\textbf{Interactions between entities and relationships.}
Recent studies have confirmed that, relationships could improve KG embeddings quality~\cite{TransE13, TransR15, NathaniCSK19}.
However, they only focus on the transfer from relationships to entities.
We describe how \textsf{TransGAT} exploits the interactions between entities and relationships to further improve the quality of KG embeddings.

In the first layer of  \textsf{TransGAT}, as mentioned earlier, the semantic information of the relationships is transmitted to form new entity embeddings $\vec{e'_{h}}$ and $\vec{e'_{t}}$ through the multi-head attention mechanism.

In the second layer of \textsf{TransGAT}, the semantic information of $\vec{e'_h}$ and $\vec{e'_{t}}$ are incorporated to generate a new relationship embedding $\vec{r'_{k}}$, as shown in Equation~(\ref{eq:transmit_e2r}).
\begin{equation}\label{eq:transmit_e2r}
\vec{r_{k}'} = \texttt{selu}(\vec{e_{h}} \otimes \vec{e_{t}}) + \vec{r_{k}}
\end{equation}
Here, $\texttt{selu}(\vec{e_{h}} \otimes \vec{e_{t}})$ denotes the offset caused by the previous training iteration of relationship embedding $\vec{r_{k}}$. Thereafter, we receive the newly generated embeddings $\vec{e_{h}'}$, $\vec{r_{k}'}$, and $\vec{e_{t}'}$, and generate the final entity embeddings according to Equation~(\ref{eq:multihead2}).
In general, the semantic information of both entities and relationships is iteratively accumulated over a $n$ layer model.
In this paper, our proposed \textsf{TransGAT} is a two-layer model (i.e., $n=2$). This is because the training cost is proportional to the number of layers while the improvement achieved by a model with more layers is rather limited.
%

\subsection{Training of TransGAT}
\label{sec:train_transGAT}


Knowledge graph embedding contains rich causalities between entities.
As mentioned before, causality can be regarded as a kind of rules to be obeyed within every triplet.
Accordingly, KG embedding models tend to use the concept of causality to design their score functions so as to train reliable embeddings. To obtain \emph{accurate} embeddings of entities and relationships, \textsf{TransGAT} obeys the causality in the knowledge graph. We start by briefly reviewing the causality proposed by the TransE model~\cite{TransR15}, from which our model is derived.

According to the idea of the seminal embedding model TransE, if a triplet $t = (e_h, r_k, e_t)$ holds, the embedding of the tail entity $e_t$ should be close to the embedding of the head entity $e_h$ plus the embedding of their relationship $r_k$, denoted by $\vec{e_h}+\vec{r_k}\approx \vec{e_t}$. Otherwise, errors may occur in the triplet.
Hence, TransE assumes the \emph{score function}
\begin{equation}\label{eq:transE}
f_r(e_h,r_k, e_t)= \left\| \vec{e_h}+\vec{r_k}-\vec{e_t} \right\|_2
\end{equation}
is low when ($e_h, r_k, e_t$) holds, and high otherwise.

\textsf{TransGAT} borrows the idea of the score function from TransE, and uses the margin-based ranking loss function for training, as defined in Equation~(\ref{eq:loss_func}).
\begin{equation}\label{eq:loss_func}
\mathcal{L}=\sum_{t \in \mathbb{D}} \sum_{t' \in \mathbb{D}'} \left[f_{r}(e_h, r_k, e_t)+ \eta - f_{r} (e_{h}', r_{k}', e_{t}')\right]_{+}
\end{equation}
Here, $\left[ x \right]_{+}$ aims to get the maximum between $x$ and $0$, $\eta > 0$ is a margin hyper-parameter, $\mathbb{D}$ denotes the set of accurate triplets sampled from the training KG $\mathcal{G}_t$, and $\mathbb{D'}$ represents the set of invalid triplets created by randomly replacing either the head or tail entity in each triplet from $\mathcal{G}_t$.

\begin{figure}[t]
\includegraphics[width=3.37in]{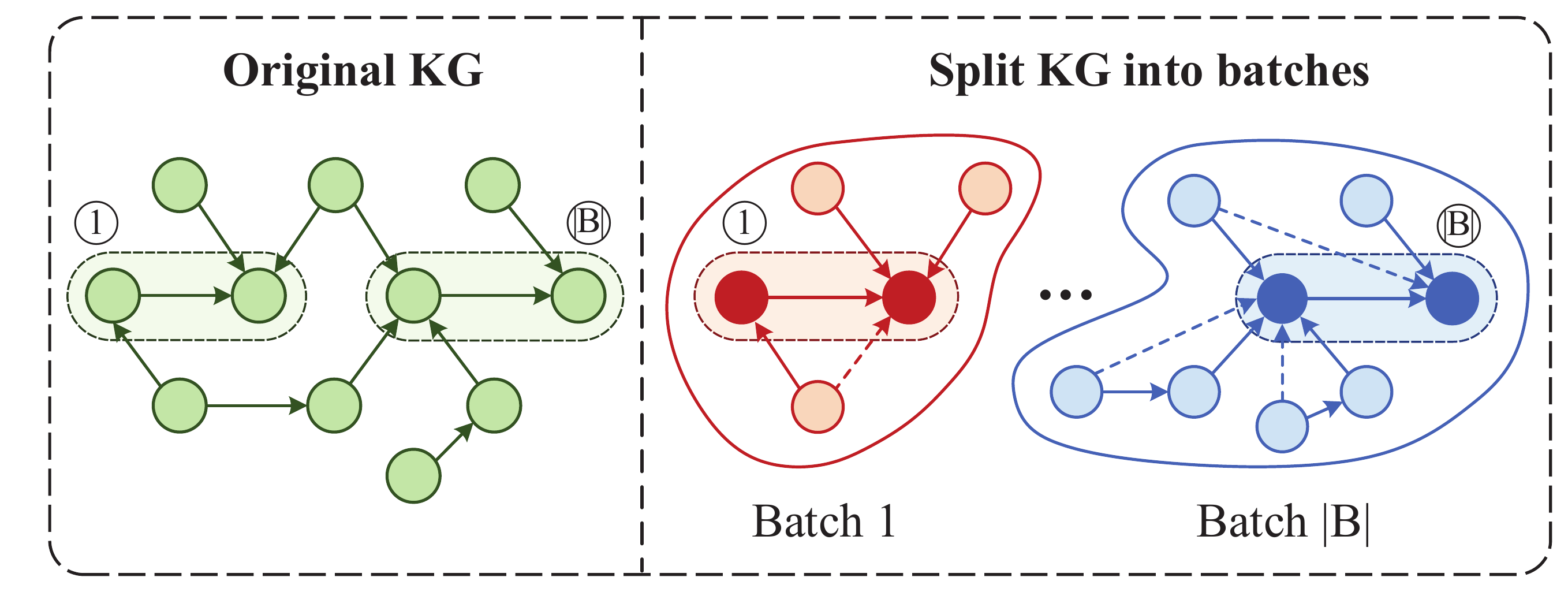}
\vspace*{-6mm}
\caption{Split a KG into multiple batches}
\label{fig:nhop}
\vspace*{-4.5mm}
\end{figure}

Note that, it is impractical to input the entire knowledge graph at once for learning embeddings.
Take the WN18 dataset, a public dataset to be used in our experimental study, as an example. It has $2,736,179$ triplets in total, including $141,442$ initial triplets and $2,594,737$ auxiliary triplets by considering 2-hop neighbors.
It will generate a $2,736,179$ (w.r.t. the triplet size) $\times$ $200$ (w.r.t. the dimensionality) triplet embedding matrix in Equation~(\ref{eq:linearTrans}). To reduce the size of matrices in the training process, we split the input knowledge graph into multiple batches, and train the data in one batch at a time. As shown in Figure~\ref{fig:nhop}, the original knowledge graph is divided into $|B|$ batches, and each batch contains one triplet and its corresponding neighbors, including both 1-hop and 2-hop neighbors. The arrows with dashed lines represent auxiliary edges generated by entities' 2-hop neighbors.

Instead of obtaining the learned embeddings directly, \textsf{TransGAT} uses ConvKB~\cite{ConvKB2018} to further optimize the global embedding properties of each triplet across every dimension. The score function of each triplet can be expressed as:
\begin{equation}\label{eq:ConvKB}
f_r(e_h, r_k, e_t)=\left(\mathop{\parallel}_{m=1}^{\Omega} \operatorname{ReLU}\left(\left[\vec{e_{h}}, \vec{r_{k}}, \vec{e_{t}}\right] * \omega^{m}\right)\right) \cdot \mathbf{W}
\end{equation}
where $\Omega$ is a hyper-parameter denoting the number of filters used, $*$ is a convolution operator, $\omega^{m}$ is the $m$-th convolutional filter, and $\mathbf{W}$ represents a linear transformation matrix. The score is high when a triplet $(e_h, r_k, e_t)$ holds, and low otherwise.

\section{data cleaning}
\label{sec:data_cleaning}
In this section, we detail the cleaning process in \textsf{KGClean}, including \emph{error detecting} and \emph{error repairing} phases.

\begin{algorithm}[t]
\label{algorithm:ALdetect}
\LinesNumbered
\DontPrintSemicolon
\caption{\textsf{AL-detect} Strategy}
    \KwIn{a training KG $\mathcal{G}_t$ and a dirty KG $\mathcal{G}_d$}
    \KwOut{the noisy part $\mathcal{G}_d^n$ and the clean part $\mathcal{G}_d^c$}
        $\mathcal{L}, \mathcal{U} \leftarrow \operatorname{\texttt{SPLIT}}(\mathcal{G}_t, \tau)$\;
        $\mathcal{G}_d^n \leftarrow \varnothing$; $\mathcal{G}_d^c \leftarrow \varnothing$\;
        train a classifier $\mathcal{C}$ based on the labeled data $\mathcal{L}$\;
        $N_{it} \leftarrow$ the maximum number of iterations\;
        $m \leftarrow 0$\;
        \While{$m < N_{it}$}
        {
            $T \leftarrow$ \texttt{queryStrategy}($\mathcal{U}$)\;
            get labels of $T$\;
            $\mathcal{L}' \leftarrow \mathcal{L} \cup T$; $\mathcal{U}' \leftarrow \mathcal{U} - T$\;
            re-train a classifier $\mathcal{C}$ based on $\mathcal{L}'$\;
            $m\leftarrow m+1$ \;
        }
        \ForEach{$t \in \mathcal{G}_d$}
        {
            $\hat{l} \leftarrow$ predict the label of $t$ according to $\mathcal{C}$\;
            \If(\!\!\tcp*[f]\!\!\!{noisy class}){$\hat{l} == -1$}
            {
                $\mathcal{G}_d^n \leftarrow$ add $t$ into $\mathcal{G}_d^n$\;
            }
            \Else(\!\!\tcp*[f]\!\!\!{clean class}){$\mathcal{G}_d^c \leftarrow$ add $t$ into $\mathcal{G}_d^c$\;}
        }
        \Return{$\mathcal{G}_d^n$ and $\mathcal{G}_d^c$}
\end{algorithm}

\subsection{Error Detecting}
\label{sec:error_detect}

Error detecting techniques either leverage quality rules \cite{FanL17} or rely on external and labeled data \cite{KATARA15} to identify erroneous values. Rule-based methods could detect errors that violate known rules. However, those methods may miss lots of errors due to the ubiquitously insufficient rules. Thus, we tend to learn an error detecting model that classifies triplets in a dirty knowledge graph into \emph{noisy} and \emph{clean} categories, according to the external labeled data. Theoretically, large-scale labeled data can benefit the classification model. Nonetheless, getting a large number of labeled data is labor-intensive. As a result, we integrate active learning (AL) techniques in learning the classification model, in order to reduce the number of labels while ensuring its reliability for detecting errors.

Algorithm \ref{algorithm:ALdetect} presents the pseudo-code of the AL-based error detecting strategy (\textsf{AL-detect} for short). Given a clean knowledge graph $\mathcal{G}_t$, \textsf{AL-detect} first splits $\mathcal{G}_t$ into a small labeled set  $\mathcal{L}$ and a large unlabeled set $\mathcal{U}$ (line 1). Then, it trains a classifier $\mathcal{C}$ based on the labeled data $\mathcal{L}$ (line 3). Next, it iteratively selects a set $T$ of data from the unlabeled pool $\mathcal{U}$ according to a \emph{query strategy} in each iteration, and adds them into the labeled set $\mathcal{L}$ after annotation to retrain the classifier $\mathcal{C}$ (lines 6-11). After $N_{it}$ iterations, a well-trained classifier $\mathcal{C}$ is generated.
\textsf{AL-detect} uses the classifier $\mathcal{C}$ to predict the label of each triplet $t$ in the dirty knowledge graph $\mathcal{G}_d$ (lines 12-13). If $t$ is noisy, it belongs to $\mathcal{G}_d^n$ (lines 14-15), otherwise, it belongs to $\mathcal{G}_d^c$ (lines 16-17).

\textsf{AL-detect} treats the \emph{query strategy} as a black box. Users have the flexibility to use any method to select data for annotation. Our current implementation uses \emph{Entropy Sampling}, which picks the triplet with the largest class prediction information entropy from the unlabeled pool, as a query strategy. Entropy Sampling is empirically found to be the optimal one in our experiments, among a series of query strategies \cite{ALselect14}, including Random Sampling, Entropy Sampling, Least Confidence, and Margin Sampling.

\textsf{AL-detect} utilizes TextCNN \cite{TextCNN14}, a popular classification model based on convolutional neural networks, as a classifier. For each triplet $t = (e_h, r_k, e_t)$, we input a $3 \times |F|$ embedding matrix (pre-trained by \textsf{TransGAT}) into the TextCNN model, where $3$ represents the number of values within the triplet, including a head entity, a relationship, and a tail entity. $F$ is the
dimensionality of each value's embedding. To facilitate the calculation, we make the dimensionality of entity embeddings equal to that of relation embeddings, denoted as $|F|=|F_e|=|F_r|$.

Then, the model outputs the probability that the triplet belongs to the clean class and the noisy class, respectively. The probability is formulated as Equation (\ref{eq:textcnn_prob}).
\begin{equation}\label{eq:textcnn_prob}
P_r(y=j|e_h, r_k, e_t) = \frac{{\rm exp}({x^\top} w_j)}{\sum_{c=1}^{C} {\rm exp}({x^\top}w_c)}
\end{equation}
where $x$ denotes a vector that is transformed from the $3 \times |F|$ embedding matrix through the TextCNN model, $w_j$ represents the weight of the triplet $t$ that belongs to the $j$-th class, and $C$ is the total number of classes.
If $P_r(y=0|e_h, r_k, e_t) > P_r(y=1|e_h, r_k, e_t)$, the triplet is considered noisy; otherwise, it is clean.
The technical details of TextCNN can be found in \cite{TextCNN14}.

As stated in Section~\ref{sec:train_transGAT}, causality is a kind of rules to be complied within each triplet.
A stronger causality indicates the triplet is more likely to be clean.
To this end, we propose a new score function $S(y=j|e_h, r_k, e_t)$ to judge the possibility that a given triplet belongs to a specific category, as defined in Equation~(\ref{eq:classify_func}). It considers not only $P_r(y=j|e_h, r_k, e_t)$, the probability output by TextCNN, but also $f_r(e_h, r_k, e_t)$, the score function of our proposed \textsf{TransGAT} (i.e., Equation~(\ref{eq:ConvKB})). We will verify that the newly proposed function $S(y=j|e_h, r_k, e_t)$ is better than the probabilistic function $P_r(y=j|e_h, r_k, e_t)$ used in the TextCNN in the experiments to be presented in Section~\ref{subsec:al-detect}.
%
\begin{equation}\label{eq:classify_func}
S(y=j|e_h, r_k, e_t) = P_r(y=j|e_h, r_k, e_t) \times f_r(e_h, r_k, e_t)
\end{equation}

\subsection{Error Repairing}

After \textsf{AL-detect} identifies noisy triplets from the knowledge graph, the next step required is to fix the errors.
We present an error repairing strategy using a novel concept of \emph{propagation power} (\textsf{PRO-repair} for short) to fix these noisy triplets. We first introduce the concept of \emph{propagation power} and then propose the \textsf{PRO-repair} strategy.

Due to the feature of the knowledge graph, that is, there are rich relationships between entities, it is unlikely that an isolated triplet exists in a KG. Given a noisy triplet, this feature guides us to pick the one with the highest probability from a large number of potential candidates. Specifically, a repaired triplet should have a great influence on its adjacent nodes of a clean KG. Meanwhile, as mentioned in Section~\ref{sec:intro}, the repaired triplet should also have a strong causality.

Therefore, we propose propagation power to represent the probability of each candidate being clean. It is reflected in two aspects, i.e., \emph{inner-power} (IP for short) and \emph{outer-power} (OP for short). Given a triplet, IP is related to the causality of the triplet, and OP represents the influence of the triplet in its neighbors. Besides, both IP and OP are relevant to the paths flowing through the triple. An explicit path on a KG provides the ability to interpret the cause of the cleaning result with the support of causal evidence (i.e., the entities and relationships contained in the path). Next, we give the formal definitions of IP and OP, and describe the paths related to them.

\noindent
\textbf{Inner-Power.} Given a triplet $t = (e_h, r_k, e_t) \in \mathcal{G}_d$, its \emph{inner-power} (IP) refers to the probability of a path that starts from $e_h$, flows through a relationship $r_k$, and reaches $e_t$, i.e., $e_{h} \stackrel{r_{k}}{\longrightarrow} e_{t}$.
In other words, IP represents the probability that $(e_h, r_k, e_t)$ holds.
The higher the probability, the more the inner-power owned by the triplet. Formally,
\begin{equation}
IP(e_h, r_k, e_t) = \operatorname{Sigmoid}(f_r(e_h, r_k, e_t))
\end{equation}
where $f_r(e_h, r_k, e_t)$ is the score function of \textsf{TransGAT} ranging from $[-\infty, +\infty]$ (defined in Equation~(\ref{eq:ConvKB})).
The \emph{Sigmoid} activation function is used to reduce the effect of outliers (i.e., maximal and minimal scores) and make scores in the range $[0, 1]$.

\noindent
\textbf{Outer-Power.} Given a triplet $t = (e_h, r_k, e_t) \in \mathcal{G}_d$, its \emph{outer-power} (OP) represents the influence of the triplet in its neighbors. The impact of OP depends on two factors, (i) the possibility that its neighbors flow into or out $t$, and (ii) the probability that $t$ and its neighbor nodes co-occur.

\begin{figure}[t]
\centering
\includegraphics[width=2.45in]{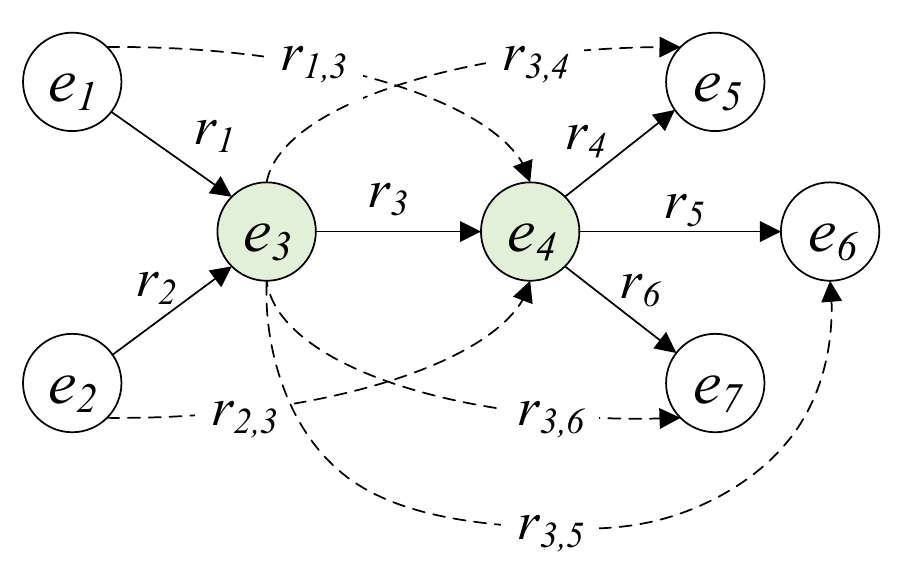}
\vspace*{-2mm}
\caption{A sample of a KG with 2-hop neighbors}
\label{fig:aux_path_example}
\vspace*{-4mm}
\end{figure}

The design of the first factor is motivated by the idea of PageRank \cite{pagerank1999}, which assigns a weight to each node that flows into $t$.
A neighbor node $e_j$ of a triplet $t = (e_h, r_k, e_t)$ refers to a node that is either connected to the head entity via an edge $e_j\stackrel{r}{\longrightarrow}e_h$ (called in-neighbor) or is reached by the tail entity via an edge $e_t\stackrel{r}{\longrightarrow}e_j$ (called out-neighbor), where $r \in \mathcal{R}$.
Accordingly, $\mu_{j}=\frac{1}{|\mathbb{I}(e_c)|}$ can be used to denote the weight of a path that links an in-neighbor $e_{j}$ to $t$ or links $t$ to an out-neighbor $e_{j}$, with $\mathbb{I}(e_c)$ representing the complete set of in-neighbors or out-neighbors of $t$. Take Figure~\ref{fig:aux_path_example}(a) as an example, and suppose $t = (e_3, r_3, e_4)$. Nodes $e_1$ and $e_2$ are the in-neighbors of $t$, and nodes $e_5$, $e_6$, and $e_7$ are the out-neighbors. Thus, $\mu_{1}=\mu_{2}=\frac{1}{2}$, and $\mu_{5}=\mu_{6}=\mu_{7}=\frac{1}{3}$. If the number of in-neighbors (or out-neighbors) is zero, the corresponding $\mu_{j} = 0$.
%

The second factor, denoted as $P_r(t, e_{j})$, represents the joint probability between $t$ and its neighbor nodes $e_{j}$.
Note that, we only consider purposely the joint probability between $t$ and its neighbors that are either directly connect to the head entity or directly reached by the tail entity, but not the nodes that are \emph{indirectly} connected to or \emph{indirectly} reached by $t$.
%
%
This is because, according to the principle of Markov Chains~\cite{markovchains1976}, $t$ is only affected by the nodes directly connected to or directly reached by it. Next, we formulate the computation of $P_r(t, e_{j})$.
In Figure~\ref{fig:aux_path_example}, $e_1$ and $e_5$ are two neighbors of $t = (e_3, r_3, e_4)$, where $e_1$ is an in-neighbor and $e_5$ is an out-neighbor. When calculating $P_r(t, e_1)$, we need to know the occurrence probability of the path $e_{1} \stackrel{r_{1}}{\longrightarrow} e_{3} \stackrel{r_{3}}{\longrightarrow} e_{4}$.
%
Since $e_1$ is a 2-hop neighbor of $e_4$, we could transform the calculation of $P_r(t, e_1)$ to the calculation of the score function $f_r(e_1, r_{1,3}, e_4)$, as stated in Equation~(\ref{eq:ConvKB}). Therefore, $r_{1,3}$ is a potential relationship learned by the \textsf{TransGAT} embedding model. Similarly, when calculating $P_r(t, e_5)$, we need to know the occurrence probability of the path $e_{3} \stackrel{r_{3}}{\longrightarrow} e_{4} \stackrel{r_{4}}{\longrightarrow} e_{5}$.
Since $e_3$ is a 2-hop neighbor of $e_5$, we can use $f_r(e_3, r_{3,4}, e_5)$ to compute $P_r(t, e_5)$.
After considering the influence of a triplet $t$ to all its neighbors, we can get its outer-power.
Formally,
\begin{equation}
OP(e_h, r_k, e_t) = \sum\limits_{\forall e_j \in \mathcal{N}_{t}} \mu_{j} \times P_r(t, e_{j})
\end{equation}
where $\mathcal{N}_{t}$ represents the set of neighbors of $t$, including both in-neighbors and out-neighbors.



By considering both the inner power and the outer power, the propagation power of an triplet $(e_h, r_k, e_t)$, denoted as $\Gamma(e_h, r_k, e_t)$, can be formulated as
\begin{equation}
\Gamma(e_h, r_k, e_t) = \frac{1}{Z}\left(IP(e_h, r_k, e_t) + OP(e_h, r_k, e_t)\right)
\end{equation}
where
$Z$ is a normalization function to make $\Gamma(e_h, r_k, e_t)$ within the interval $[0, 1]$.

\begin{algorithm}[t]
\label{algorithm:PRO-repair}
\LinesNumbered
\DontPrintSemicolon
\caption{\textsf{PRO-repair} Strategy}
    \KwIn{a noisy triplets set $\mathcal{G}_d^n$, a clean triplets set $\mathcal{G}_d^c$}
    \KwOut{the repaired triplets set $\mathcal{G}_d^{n*}$}\
    $\mathcal{G}_d^{n*} \leftarrow \varnothing$\;
    \ForEach{$t \in \mathcal{G}_d^n$}
    {
        $\mathbb{L}_h, \mathbb{L}_r, \mathbb{L}_t \leftarrow$ \texttt{candidatesRank}($t, k$)\;
        \ForEach{$e_h^{*} \in \mathbb{L}_h$}
        {
            compute $\Gamma(e_h^{*}, r_k, e_t)$\;
        }
        \ForEach{$r_k^{*} \in \mathbb{L}_r$}
        {
            compute $\Gamma(e_h, r_k^{*}, e_t)$\;
        }
        \ForEach{$e_t^{*} \in \mathbb{L}_t$}
        {
            compute $\Gamma(e_h, r_k, e_t^{*})$\;
        }
        $\hat t \leftarrow$ get the triplet with maximum $\Gamma$\;
        $t \gets \hat t$\;
        $\mathcal{G}_d^{n*} \leftarrow$ add $\hat t$ into $\mathcal{G}_d^{n*}$\;
    }
    \Return{$\mathcal{G}_d^{n*}$}
\end{algorithm}

After presenting the concept of propagation power, we are ready to introduce the details of the \textsf{PRO-repair} strategy, whose pseudo-code is presented in Algorithm \ref{algorithm:PRO-repair}. According to the minimality principle of repair cost (i.e., minimizing the impact on the dataset by trying to preserve as many values as possible) \cite{KATARA15}, we assume that there is one and only one error value in each triplet $t_i \in \mathcal{G}_d^n$. In the following, we enumerate the only three scenarios where errors may occur.
For each triplet $(e_h, r_k, e_t) \in \mathcal{G}_d^n$, (i) if $e_h$ is erroneous, we generate candidate triplets, denoted as $(e_h^{*}, r_k, e_t)$, by replacing $e_h$ with another entity $e_h^{*} \in \mathcal{E}$.
(ii) If $r_k$ is dirty, we generate candidate triplets, denoted as $(e_h, r_k^{*}, e_t)$, by replacing $r_k$ with another relationship $r_k^{*} \in \mathcal{R}$ (lines 6-7). %
(iii) If $e_t$ is erroneous, we generate candidate triplets, denoted as $(e_h, r_k, e_t^{*})$, by replacing $e_t$ with another entity $e_t^{*} \in \mathcal{E}$ (lines 8-9). Note that, we purposely consider the original triplet $t_i$ also as a candidate to minimize the impact of misclassification (e.g., a clean triplet is classified as noisy by \textsf{AL-detect}), and we expect the propagation power of the clean triplet to be higher than other candidates.

After getting all candidates for repairing the triplet $(e_h$, $r_k$, $e_t)$, we compute their respective propagation power and then pick the one with the highest propagation power as the ultimate repair decision (lines 10-11).
Since walking through the \textsf{PRO-repair} strategy on all candidates of all noisy triplets is expensive, we choose the top-$k$ candidates to compute their propagation power for each noisy triplet $t_i$.
Specifically, we calculate the inner-power of each candidate, rank them based on descending order, and prune the candidates that have low inner-power (line 3).
This is because, given a triplet $t = (e_h^{*}, r_k, e_t)$, if the probability that $(e_h^{*}, r_k, e_t)$ holds is low, the triplet $t$ is unlikely to be a candidate.
Our current implementation only leaves the top-10 ($k=10$) candidates, and removes other candidates. The experimental results in Section \ref{sec:analyse_transGAT} confirm that the top-10 candidates are sufficient to include the vast majority of the optimal ones for repairing noisy triplets with significantly reduced cost. The algorithm stops when all noisy triplets are repaired (line 13).

\section{experiments}
\label{sec:experiments}

In this section, we present a comprehensive experimental evaluation. In what follows, we first evaluate the performance of each single component of our newly proposed data cleaning framework \textsf{KGClean}, including the novel knowledge graph embedding model \textsf{TransGAT}, the active learning enabled error detection algorithm \textsf{AL-detect}, and the error repairing strategy \textsf{PRO-repair} based on propagation power; and then, we report the performance of \textsf{KGClean} in terms of identifying and correcting the errors in a knowledge graph.
%

\subsection{Experimental Setup}

\noindent
\textbf{Datasets.}
In the experiments, we use four typical knowledge graphs:
(i) \emph{Alyawarra Kinship} \cite{LinSX18} containing kinship relationships among members of the Alyawarra tribe from Central Australia, in total of 104 entities and 25 types of relationships;
(ii) \emph{UMLS} \cite{KokD07} including data from the Unified Medical Language System, a biomedical ontology, in total of 135 entities and 46 types of relationships;
(iii) \emph{WN18} \cite{TransE13}, a subset of WordNet with 40,943 entities and 18 types of relationships; and
(iv) \emph{WN18RR} \cite{Conv2D18}, another subset of WordNet without inverse relationships, with 40,943 entities and 11 types of relationships.
A relationship $r$ between two entities $e_1$ and $e_2$ is considered \emph{invertible}, iff $e_1 \stackrel{r}{\longrightarrow} e_2$ $\Rightarrow$ $e_2 \stackrel{r'}{\longrightarrow} e_1$, where $r'$ is the inverse relationship of $r$. Accordingly, the embedding of these relationships can be easily learned. Previous studies \cite{Conv2D18} find that WN18 contains \emph{inverse relationships}, whereby one can achieve state-of-the-art results using the features of inverse relationships. That is the reason we also include the dataset WN18RR that has removed all the inverse relationships. Table \ref{tab:datasets} lists the detailed statistics of these datasets.

\begin{table}[t]\scriptsize
\caption{Statistics of Datasets Used in Experiments}
\label{tab:datasets}
\begin{tabular}{|p{1.05cm}|p{1.15cm}|p{1.95cm}|p{.7cm}p{.5cm}p{.65cm}|}
\hline
\multirow{2}{*}{\textbf{Dataset}} & \multirow{2}{*}{\textbf{\#Entities}} & \multirow{2}{*}{\textbf{\#Relationships}} & \multicolumn{3}{c|}{\textbf{\#Edges}} \\ \cline{4-6}
                         &                             &                              & Train    & Valid   & Test    \\ \hline
Kinship                  & 104                         & 25                           & 8,544    & 1,068   & 1,074   \\
UMLS                  & 135                         & 46                           & 5,216    & 652   & 661   \\
WN18                     & 40,943                      & 18                           & 141,442  & 5,000   & 5,000   \\
WN18RR                   & 40,943                      & 11                           & 86,835   & 3,034   & 3,134   \\ \hline
\end{tabular}
\vspace*{-4mm}
\end{table}


Each of these four public datasets has been split into three disjoint sets: the training set, the validation set, and the test set. In the knowledge representations phase, (i) the training set is used to learn the embeddings of entities and relationships; (ii) the validation set is utilized to estimate the embeddings' accuracy of each epoch in the training process; and (iii) the test set is employed to evaluate the performance of embeddings. In the data cleaning phase, errors are artificially introduced by randomly replacing the values of entities or relationships in the training set and test set.
(i) The training set is used to train the classification model. To ensure the balanced distribution of noisy and clean triplets, we randomly add errors to 50\% of triplets in the training set, and the remaining triplets are labeled as clean. Then, we randomly select 5\% triplets from the entire training set as initial labels, and the leftover triplets are treated as unlabeled data in order to simulate the process of active learning.
(ii) The test set is used as a dirty KG that needs to be cleaned, where the number of noisy triplets injected is controlled by the parameter \emph{error rate}.

\noindent
\textbf{Competing Methods.} To verify the performance of our proposed KG embedding model \textsf{TransGAT} in performing the cleaning task, we compare it against two competing KG embedding models\footnote{We download publicly available source codes to reproduce results of the competing KG embedding models on all the datasets.}:
\begin{itemize}\setlength{\itemsep}{-\itemsep}
  \item \textbf{TransE} \cite{TransE13}: As the seminal work for translation-based model, TransE first projects the values of entities and relationships onto a low-dimension vector space as $\bm{e_h}$, $\bm{r}$, $\bm{e_t}$ $\in \mathbb{R}^{k}$,  and then, it translates the semantics from head entities to tail entities by relationships, which requires $\vec{e_h}+\vec{r_k} \approx \vec{e_t}$ when triplet $(e_h, r_k, e_t)$ holds.
  \item \textbf{KBGAT} \cite{NathaniCSK19}: KBGAT is the state-of-the-art knowledge graph embedding model that generalizes and extends graph attention mechanisms to capture both entity and relationship features in a multi-hop neighborhood of each given entity.
\end{itemize}

\noindent
\textbf{Implementation Details:}
Our \textsf{KGClean} is implemented in Python 3.6 on Pytorch 1.1.
The experiments were conducted on an Intel(R) Xeon(R) Silver 4110 2.10GHz processors (8 physical cores and 32 CPU threads) with 128GB RAM accelerated by a NVIDIA GeForce RTX 2080 Ti GPU.


\begin{table*}[t]
\caption{The performance of \textsf{TransGAT}, TransE, and KBGAT on Kinship, WN18, WN18RR, and FB15K237. Hits@N values are in percentage. The best scores are in bold.}
\label{tab:transGAT_results}
\setlength{\tabcolsep}{4mm}{
\begin{tabular}{|l|ccccc||ccccc|}
\hline
                                   & \multicolumn{5}{c||}{\textbf{UMLS}}                                                                                                                                  & \multicolumn{5}{c|}{\textbf{Kinship}}                                                                                                                               \\ \cline{2-11}
                                   & MR                             & MRR                             & Hits@1                         & Hits@3                         & Hits@10                        & MR                             & MRR                             & Hits@1                         & Hits@3                         & Hits@10                        \\ \hline
TransE                             & 1.77                           & 0.797                           & 64.1                           & 92.1                           & 99.2                           & 6.80                           & 0.309                           & 0.9                            & 64.3                           & 84.1                           \\
KBGAT                              & \textbf{1.11} & \textbf{0.990}                           & \textbf{98.6}                           & \textbf{99.5}                           & \textbf{99.8} & 1.94                           & 0.904                           & 85.9                           & 94.1                           & \textbf{98.0}                           \\ \hline
\textbf{TransGAT} & \textbf{1.11}                           & \textbf{0.990} & \textbf{98.6} & \textbf{99.5} & \textbf{99.8}                           & \textbf{1.84} & \textbf{0.940} & \textbf{91.7} & \textbf{94.7} & 97.9 \\ \hline\hline
                                   & \multicolumn{5}{c||}{\textbf{WN18}}                                                                                                                                           & \multicolumn{5}{c|}{\textbf{WN18RR}}                                                                                                                                         \\ \cline{2-11}
                                   & MR                             & MRR                             & Hits@1                         & Hits@3                         & Hits@10                        & MR                             & MRR                             & Hits@1                         & Hits@3                         & Hits@10                        \\ \hline
TransE                             & \textbf{158}  & 0.768                           & 61.5                           & 92.3                           & 95.5                           & 2300                           & 0.279                           & 4.3                            & 44.1                           & 53.2                           \\
KBGAT                              & 213                            & 0.496                           & 23.6                           & 72.9                           & 91.6                           & 1940                           & 0.440                           & 36.1                           & 48.3                           & 58.1                           \\ \hline
\textbf{TransGAT} & 159                            & \textbf{0.890} & \textbf{84.1} & \textbf{93.5} & \textbf{95.8} & \textbf{1928} & \textbf{0.450} & \textbf{37.5} & \textbf{49.1} & \textbf{58.2} \\ \hline
\end{tabular}}
\vspace*{-4mm}
\end{table*}

\subsection{Results on TransGAT}
\label{sec:analyse_transGAT}
\textsf{KGClean} proposes a new KG embedding model, namely \textsf{TransGAT}, with the objective to preserve rich and accurate semantic information related to entities and relationships of a given KG. In order to verify the effectiveness of \textsf{TransGAT}, we adopt \emph{link prediction} to evaluate the performance of our proposed \textsf{TransGAT}, as compared with two competing KG embedding models TransE and KBGAT.
%
Link prediction is a common protocol for knowledge graph evaluation. It aims to predict head/tail entity (represented by ``?'') that is missing in a specified triplet, which is in the form of (?, $r_k$, $e_t$) or ($e_h$, $r_k$, ?).
Following previous work~\cite{NathaniCSK19}, we select 60\% of triplets in the test set, and randomly delete either the head or the tail entity from those triplets. Then, we predict the missing entities via generating the candidate triplets. We also assign scores to the candidates using the score functions introduced by respective KG embedding models (e.g., Equation~(\ref{eq:ConvKB}) by \textsf{TransGAT}, and Equation~(\ref{eq:transE}) by TransE). Subsequently, we rank all the candidates (including the correct triplet whose value matches the group truth) based on ascending order of the scores, and find the rank of the correct triplets (which shall be ranked the first ideally).
%
We then report three popular metrics to evaluate the quality of the score functions adopted by different KG embedding models, i.e., \emph{MeanRank (MR)} that reports the mean rank of the correct triplet for each link prediction task (the smaller, the better); \emph{Mean Reciprocal Rank (MRR)} of a link prediction result that is multiplicative inverse of the rank of the correct answer (i.e., $\frac{1}{rank_{t}}$, the larger, the better), and \emph{Hits@N} which is the proportion of correct triplets in the top $N$ ranks for $N = 1$, $3$, and $10$ (the higher, the better).

\begin{figure*}[t]
\centering
\includegraphics[width=0.7\textwidth]{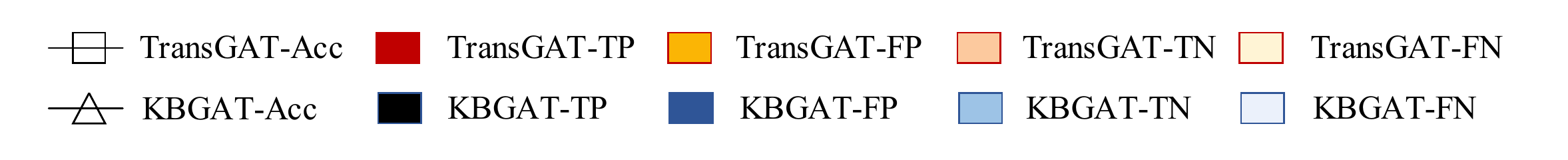}\\
\subfigure[\emph{UMLS}]{
   \includegraphics[width=1.75in]{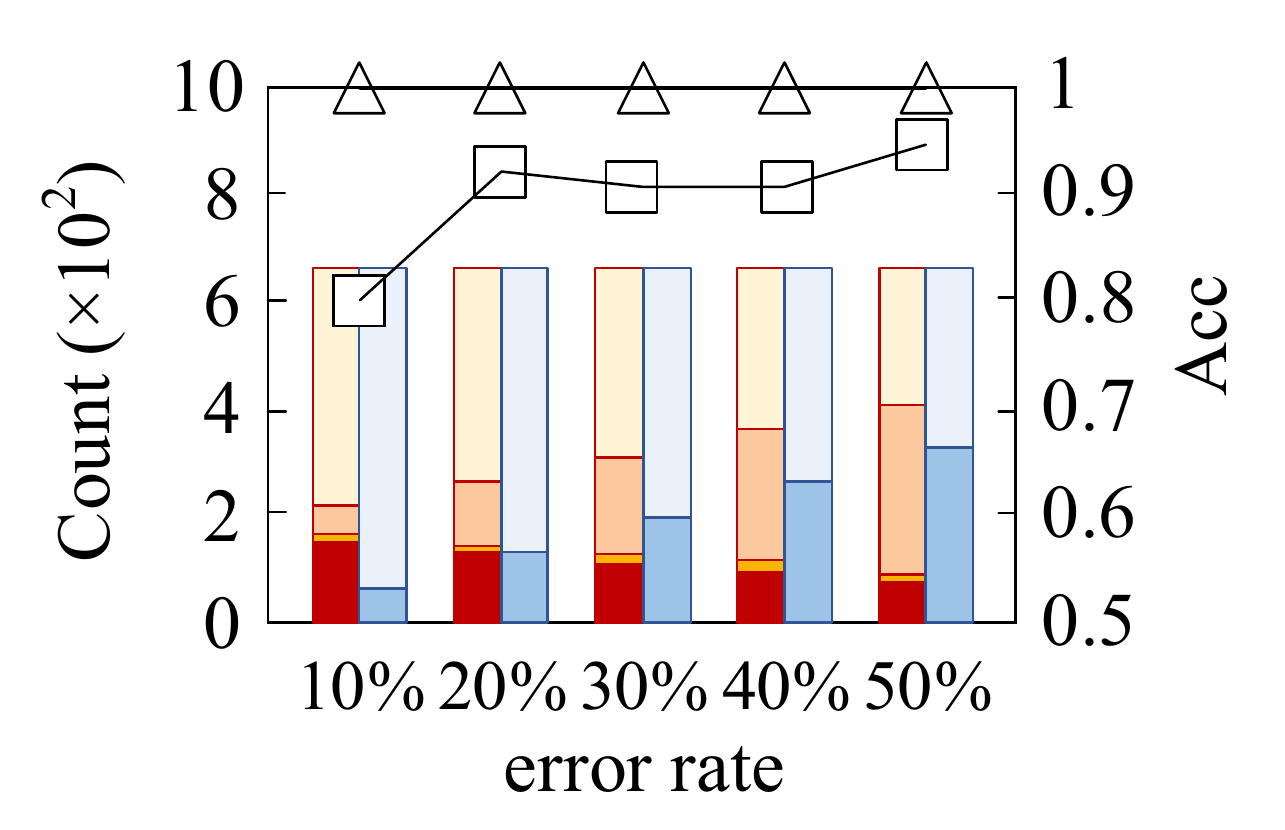}
  }\hspace*{-3mm}
\subfigure[\emph{Kinship}]{
   \includegraphics[width=1.75in]{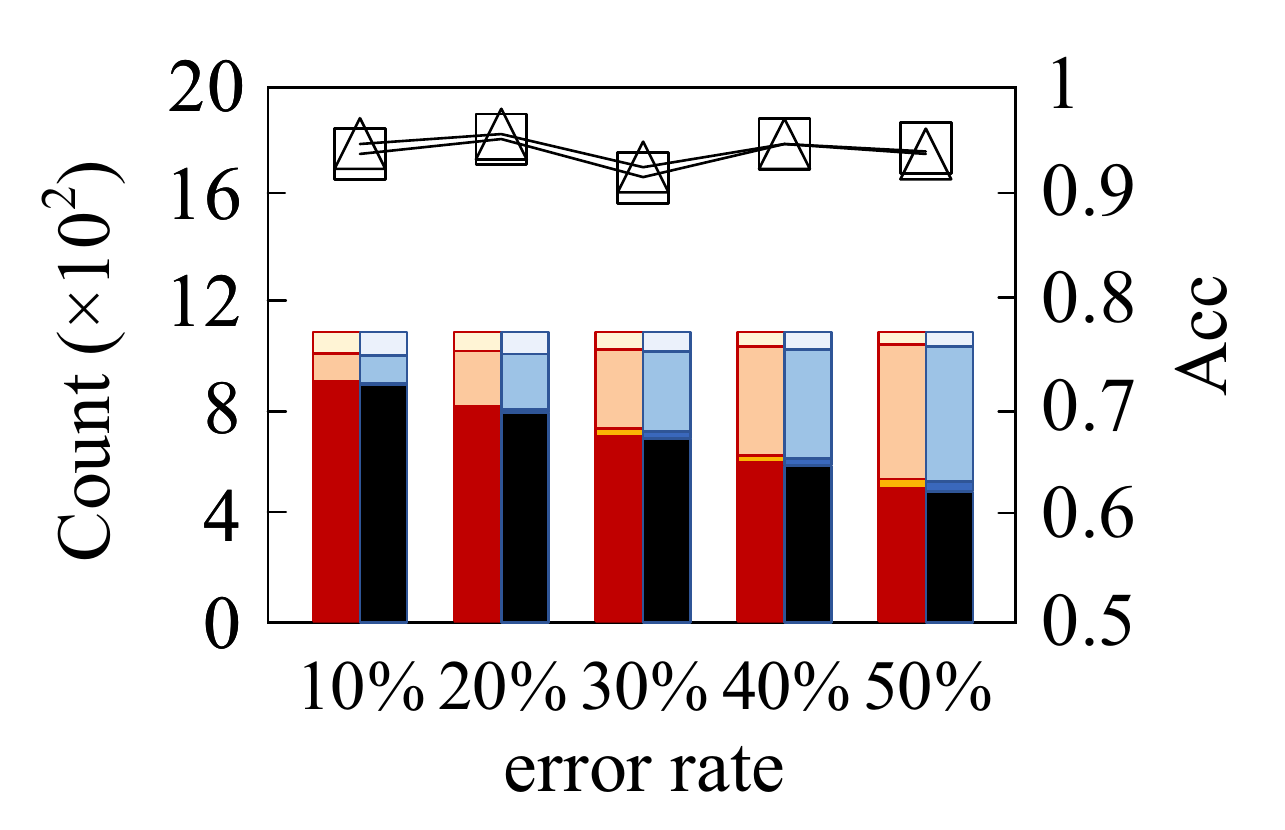}
  }\hspace*{-3mm}
  \subfigure[\emph{WN18}]{
   \includegraphics[width=1.75in]{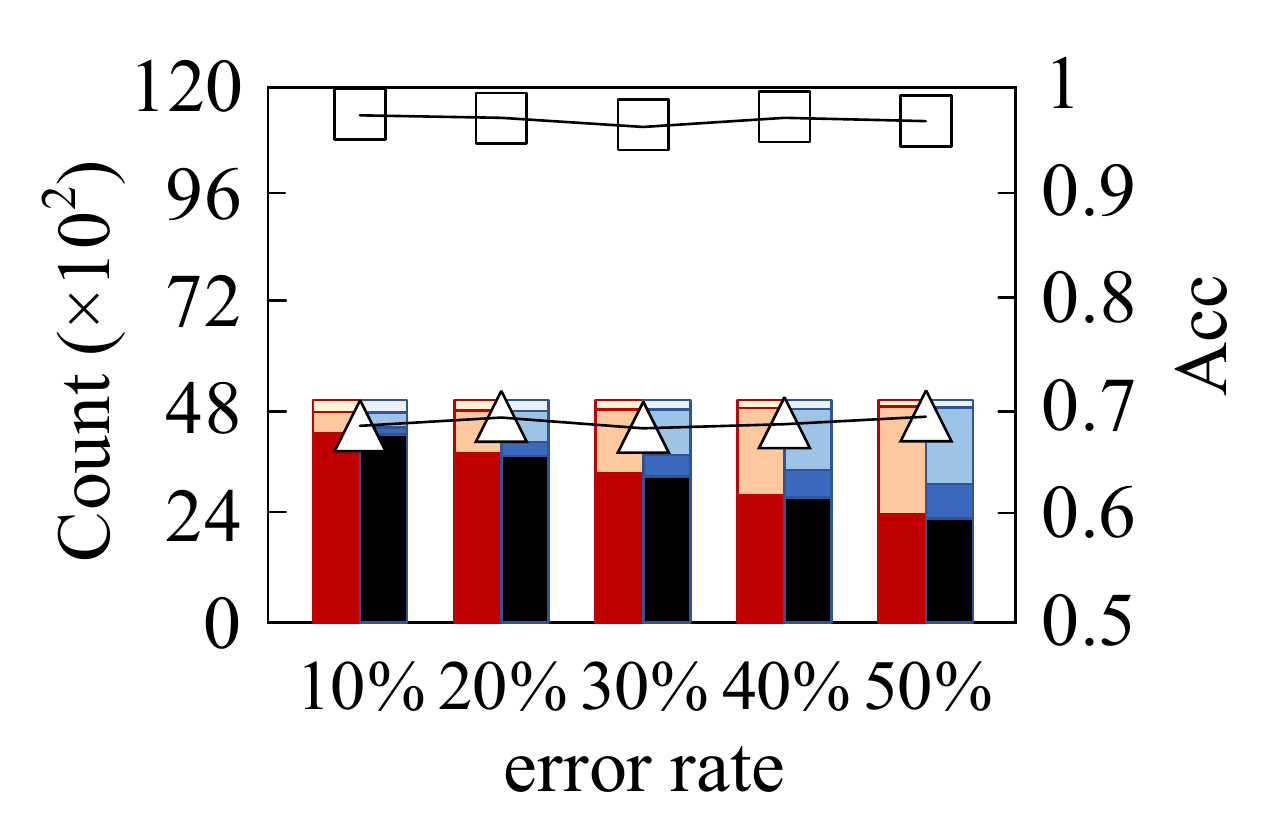}
  }\hspace*{-3mm}
\subfigure[\emph{WN18RR}]{
   \includegraphics[width=1.75in]{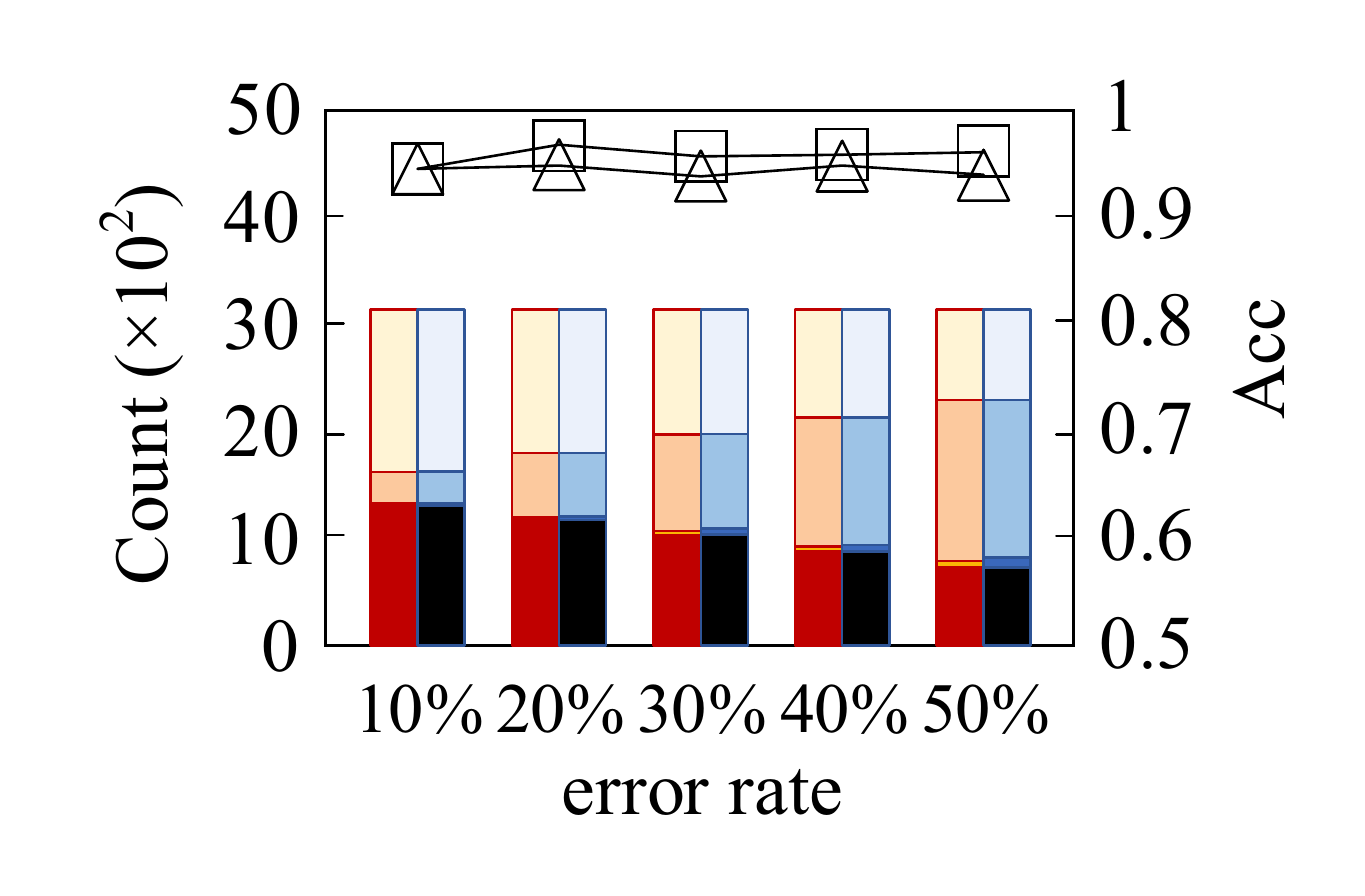}
  }
\vspace*{-4mm}
\caption{The performance of \textsf{AL-detect} vs. error rate}
\label{fig:al-detect-error-rate}
\vspace*{-2mm}
\end{figure*}

\begin{figure*}[t]
\centering
\includegraphics[width=0.38\textwidth]{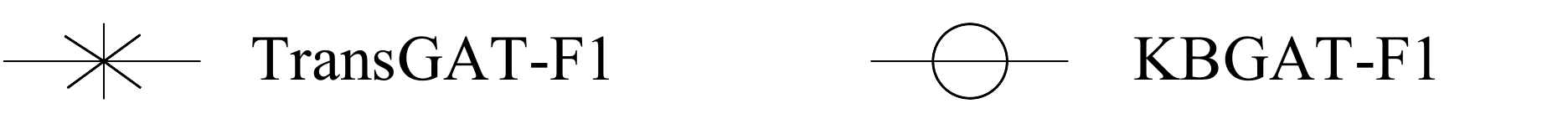}\\
\hspace*{-2mm}
\subfigure[\emph{UMLS}]{
   \includegraphics[width=1.8in]{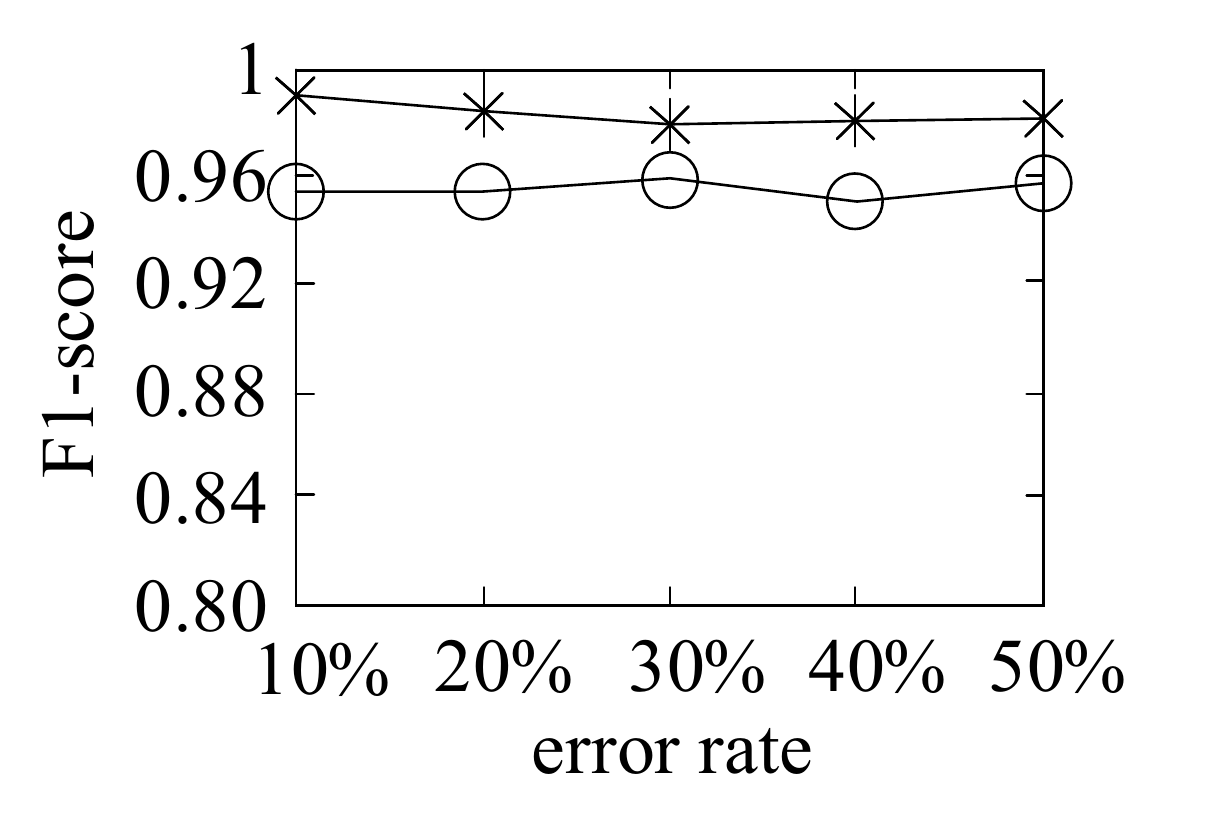}
  }\hspace*{-4mm}
\subfigure[\emph{Kinship}]{
   \includegraphics[width=1.8in]{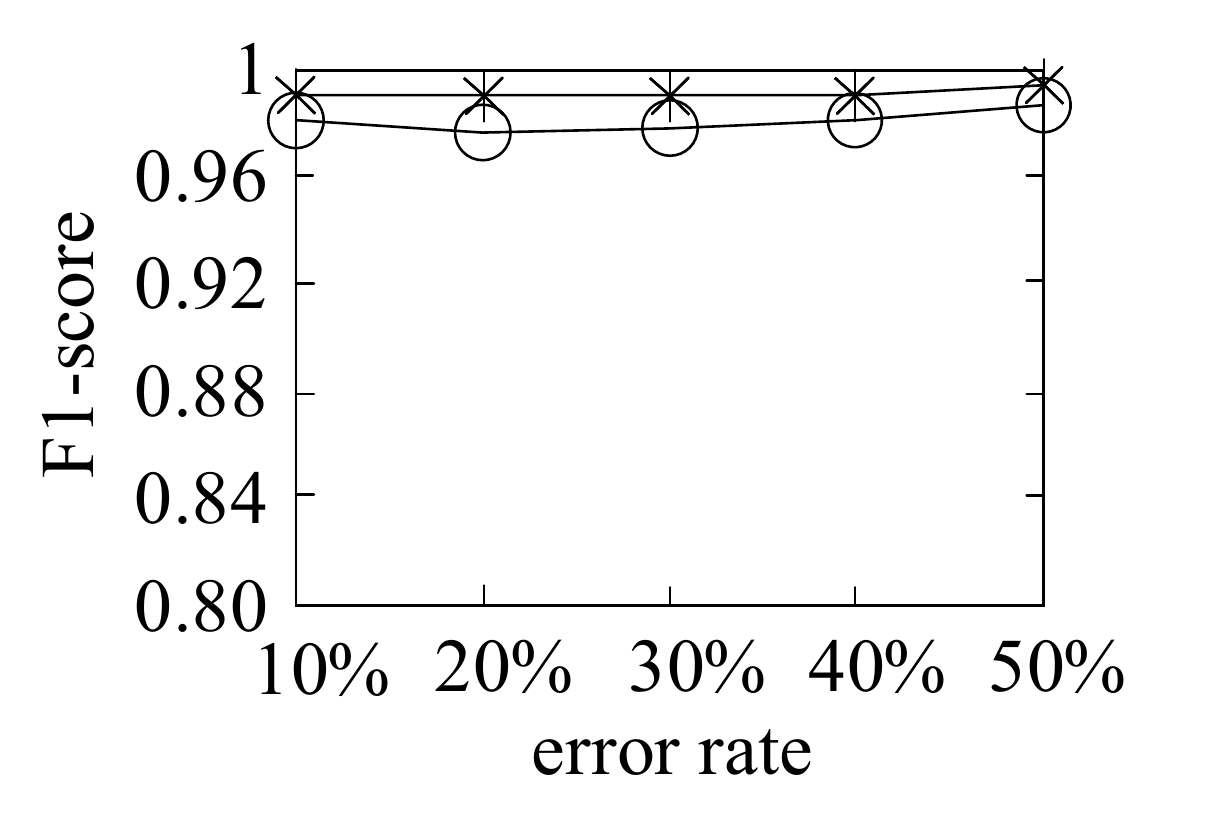}
  }\hspace*{-4mm}
  \subfigure[\emph{WN18}]{
   \includegraphics[width=1.8in]{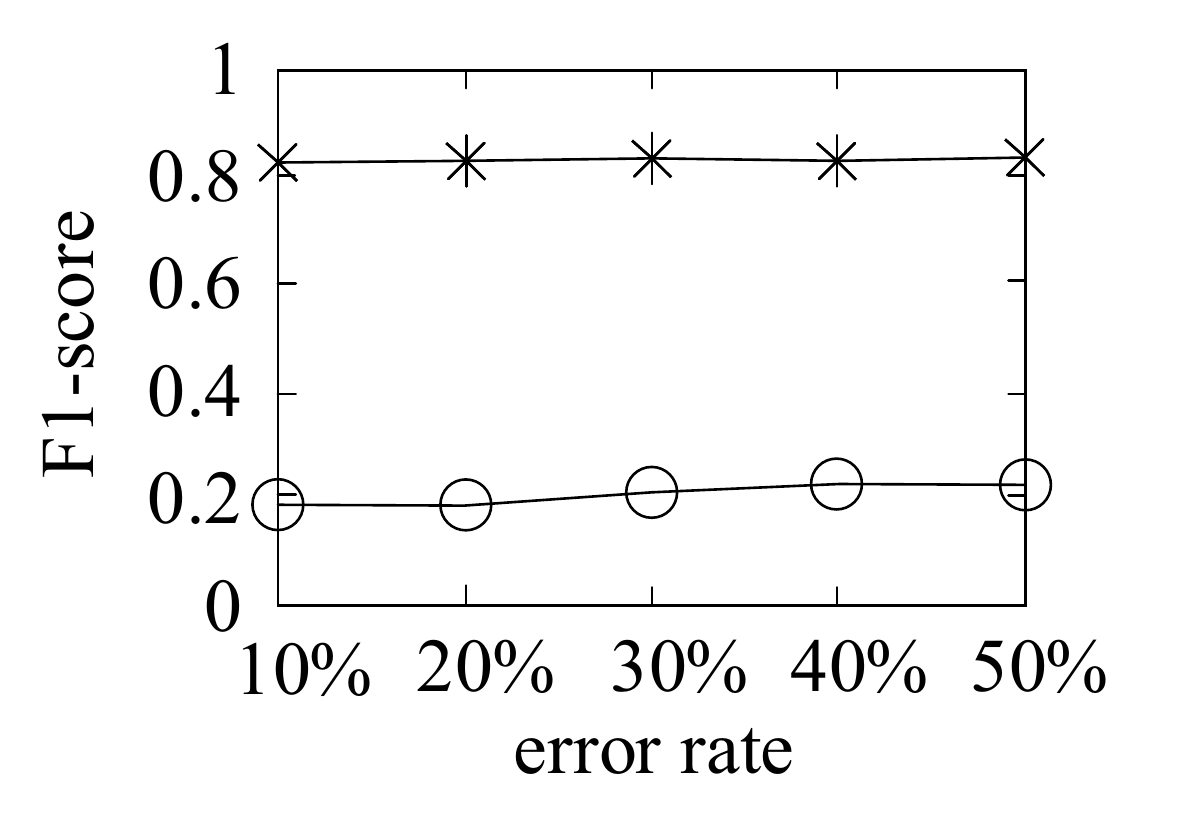}
  }\hspace*{-4mm}
\subfigure[\emph{WN18RR}]{
   \includegraphics[width=1.8in]{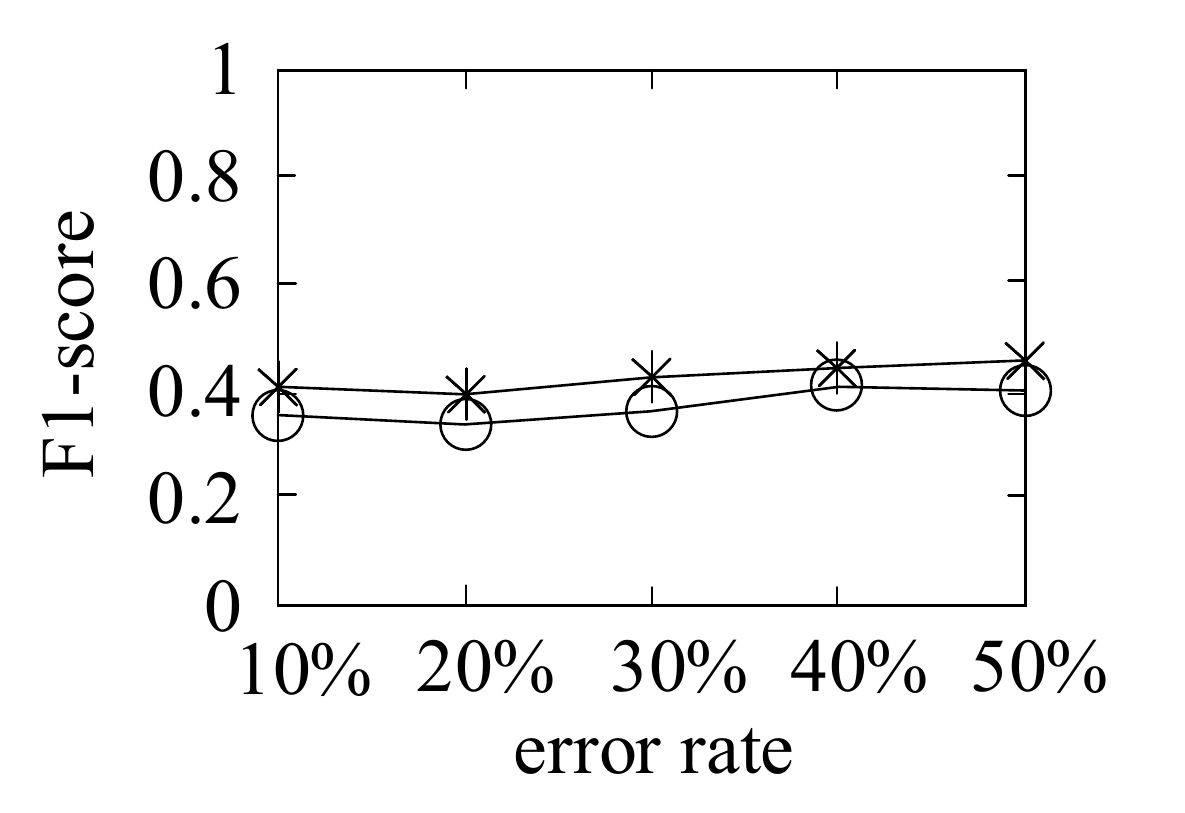}
  }
\vspace*{-4mm}
\caption{The performance of \textsf{PRO-repair} vs. error rate}
\label{fig:pro-repair-error-rate}
\vspace*{-2mm}
\end{figure*}

Table \ref{tab:transGAT_results} shows the results of link prediction. It is observed that our proposed \textsf{TransGAT} outperforms the other two models in most of the cases.
One exception is UMLS, where the results of KBGAT are comparable to that of \textsf{TransGAT}.
%
The reason is that, the relationships that connect different entities in this dataset are quite diverse. The extra information about interactions between entities and relationships is not very useful.
On the contrary, \textsf{TransGAT} performs consistently better than KBGAT in the other three datasets, which demonstrates that considering the interactions between entities and relationships can improve the precision of link prediction.
%
The second observation is that the embeddings of \textsf{TransGAT} are more accurate than KBGAT for the Hits@1 metric.
Since the cleaning tasks in \textsf{KGClean} are related to the top candidates, higher values of Hit@1 ensure that the \textsf{TransGAT}-based cleaning results are superior to the KBGAT-based cleaning results. In the remaining experiments, we only adopt \textsf{TransGAT} and KBGAT as the KG embedding models but ignore TransE because it is not comparable to the other two.

\subsection{Results on AL-detect}
\label{subsec:al-detect}

Second, we evaluate the performance of \textsf{AL-detect}, which
adopts the classification approaches to separate noisy triplets from the clean ones. To better evaluate the accuracy of the classification results, we utilize four metrics: (i) \emph{true-positive} (TP) represents the number of truly detected clean triplets; (ii) \emph{true-negative} (TN) denotes the number of truly detected noisy triplets; (iii) \emph{false-positive} (FP) refers to the number of missed erroneous triplets; and (iv) \emph{false-negative} (FN) is the number of clean triplets that are detected as errors.
A new metric \emph{TN-accuracy}, denoted as Acc, is introduced to verify the quality of \textsf{AL-detect}. TN-accuracy is defined as the fraction of truly detected erroneous triplets over the total number of the erroneous triplets, denoted as $\frac{TN}{TN+FP}$. We do not consider the FN metric for evaluation, since we empirically found that FN has little impact on the cleaning results of \textsf{KGClean}. As reported in Section \ref{sec:analyse_transGAT}, the group truth triplet, also considered as a candidate, is expected to have the highest score. We implement two versions of \textsf{AL-detect}, one trained by \textsf{TransGAT} and the other trained by KBGAT.

\vspace*{0.1in}
\noindent
\textbf{Varying Error Rate.}
We study the TN-accuracy of the \textsf{AL-detect} strategy by varying the error rate from 10\% to 50\%.
Figure \ref{fig:al-detect-error-rate} shows the corresponding results. The first observation is that \textsf{AL-detect} performs stably in terms of TN-accuracy (with minor disturbances) when the error rate increases. This is because the classification model of \textsf{AL-detect} learns the semantic information contained within the embeddings, and it is able to distinguish erroneous triplets from the clean ones. In other words, its performance is independent of the error rate. One exception is the UMLS dataset, where the TN-accuracy of the \textsf{TransGAT}-based \textsf{AL-detect} jumps from initial 80\% to later 90\% when the error rate increases from 10\% to larger values.
%
The reason is that UMLS is a small dataset which contains only 66 truly erroneous triplets. Consequently, even a small number of wrongly predicted triplets may cause a considerable drop in the accuracy.
The second observation is that \textsf{TransGAT}-based \textsf{AL-detect} achieves higher F1-scores than KBGAT-based \textsf{AL-detect}, especially in WN18 dataset, contributed by the superior capacity of \textsf{TransGAT} to accurately preserve the semantic information of triplets.
%
However, there is also an exception. In UMLS dataset, KBGAT-based \textsf{AL-detect} outperforms \textsf{TransGAT}-based \textsf{AL-detect}. The reason is that KBGAT-based \textsf{AL-detect} predicts all triplets as errors, including all clean triplets and all noisy triplets, while UMLS dataset does have a larger number of noisy triplets. It reflects the unreliable classification results when using KBGAT embeddings and the fact that KBGAT embeddings are biased towards the noisy triplets.

\vspace*{0.1in}
\noindent
\textbf{Model Variants.}
We validate the importance of using the score function of \textsf{TransGAT} for \textsf{AL-detect} (i.e., Equation~(\ref{eq:classify_func})). To demonstrate the advantage of Equation~(\ref{eq:classify_func}), we implement two versions of \textsf{AL-detect}, the one based on TextCNN and the one based on \textsf{TransGAT}. We want to highlight that \textsf{AL-detect} utilizes TextCNN but it further improves the classification power by considering not only the probability function (i.e., Equation~(\ref{eq:textcnn_prob})) but also the score of each triplet in \textsf{TransGAT}.
Since the previous experiments show that the performance of \textsf{AL-detect} is independent of the error rate, we perform this experiment with a 10\% error rate. The results are reported in Table~\ref{tab:model_variants}. Although TextCNN based strategy achieves higher accuracy, we could observe that TextCNN classifies \emph{all} the triplets as noisy, which could guarantee 100\% accuracy as its FP value is always zero. This strategy is able to detect truly erroneous triplets but it also results in a larger number of FNs. We want to emphasize that this classification results are abysmal, because the classifier does not seem to work.
All these FN triplets will become the input to the subsequent error repairing phase, which significantly increases the cost of error repairing phase. In contrast, \textsf{TransGAT} based strategy provides a much better solution to resolve the imbalance. Its scoring function is effective in terms of distinguishing the noise triplets from the clean ones. Consequently, it is able to efficiently reduce the number of FN triplets for the subsequent repairing phase without suffering much from the accuracy loss of classification.

\begin{table}[!]
\caption{The performance of \textsf{AL-detect} vs. model variants}
\label{tab:model_variants}
\setlength{\tabcolsep}{3mm}{
\begin{tabular}{|c|c|cccc|c|}
\hline
\textbf{Dataset}         & \textbf{Function} & \textbf{TP}   & \textbf{FP} & \textbf{TN}  & \textbf{FN}   & \textbf{Acc}   \\ \hline
\multirow{2}{*}{UMLS}    & Eq.~(\ref{eq:textcnn_prob}) & 5    & 0  & 66  & 590  & 1     \\ \cline{2-7}
                         & Eq.~(\ref{eq:classify_func}) & 152  & 13 & 53  & 443  & 0.80 \\ \hline
\multirow{2}{*}{Kinship} & Eq.~(\ref{eq:textcnn_prob}) & 7    & 0  & 107 & 960  & 1     \\ \cline{2-7}
                         & Eq.~(\ref{eq:classify_func}) & 887  & 7  & 100 & 80   & 0.93 \\ \hline
\multirow{2}{*}{WN18}    & Eq.~(\ref{eq:textcnn_prob}) & 0    & 0  & 500 & 4500 & 1     \\ \cline{2-7}
                         & Eq.~(\ref{eq:classify_func}) & 4238 & 13 & 487 & 262  & 0.97 \\ \hline
\multirow{2}{*}{WN18RR}  & Eq.~(\ref{eq:textcnn_prob}) & 0    & 0  & 313 & 2821 & 1     \\ \cline{2-7}
                         & Eq.~(\ref{eq:classify_func}) & 1310 & 17 & 296 & 1511 & 0.95 \\ \hline
\end{tabular}}
\vspace*{-4mm}
\end{table}

\subsection{Results on PRO-repair}

Next, we investigate the performance of \textsf{PRO-repair} and its sensitivity to different parameters, including KG embedding models employed and the error rate.
%


\begin{figure*}[!]
\centering
\includegraphics[width=0.5\textwidth]{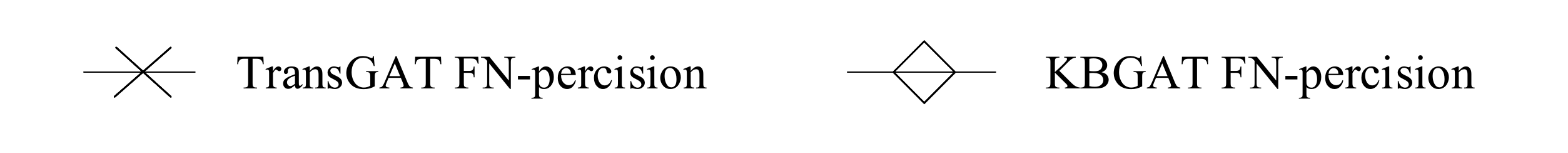}\\
\vspace*{-2.5mm}
\hspace*{-2mm}
\subfigure[\emph{UMLS}]{
   \includegraphics[width=1.8in]{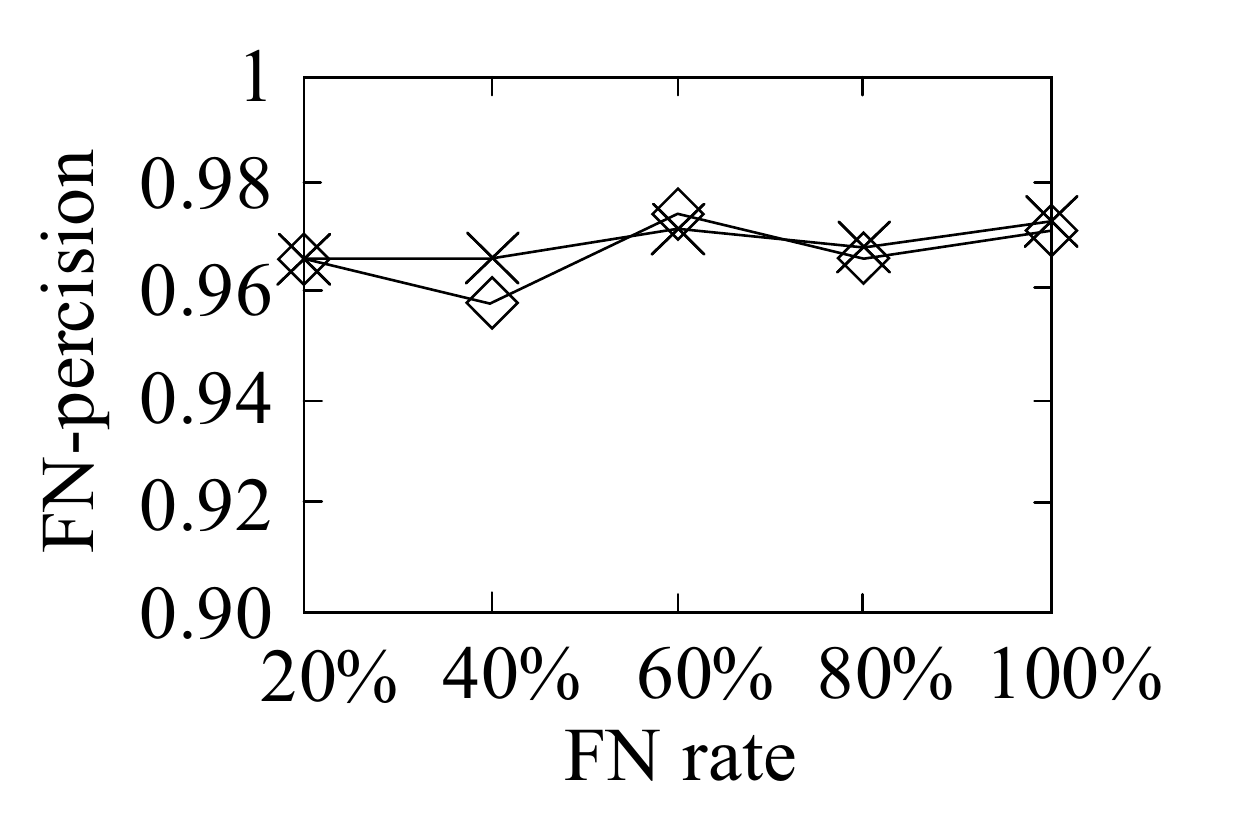}
  }\hspace*{-4mm}
\subfigure[\emph{Kinship}]{
   \includegraphics[width=1.8in]{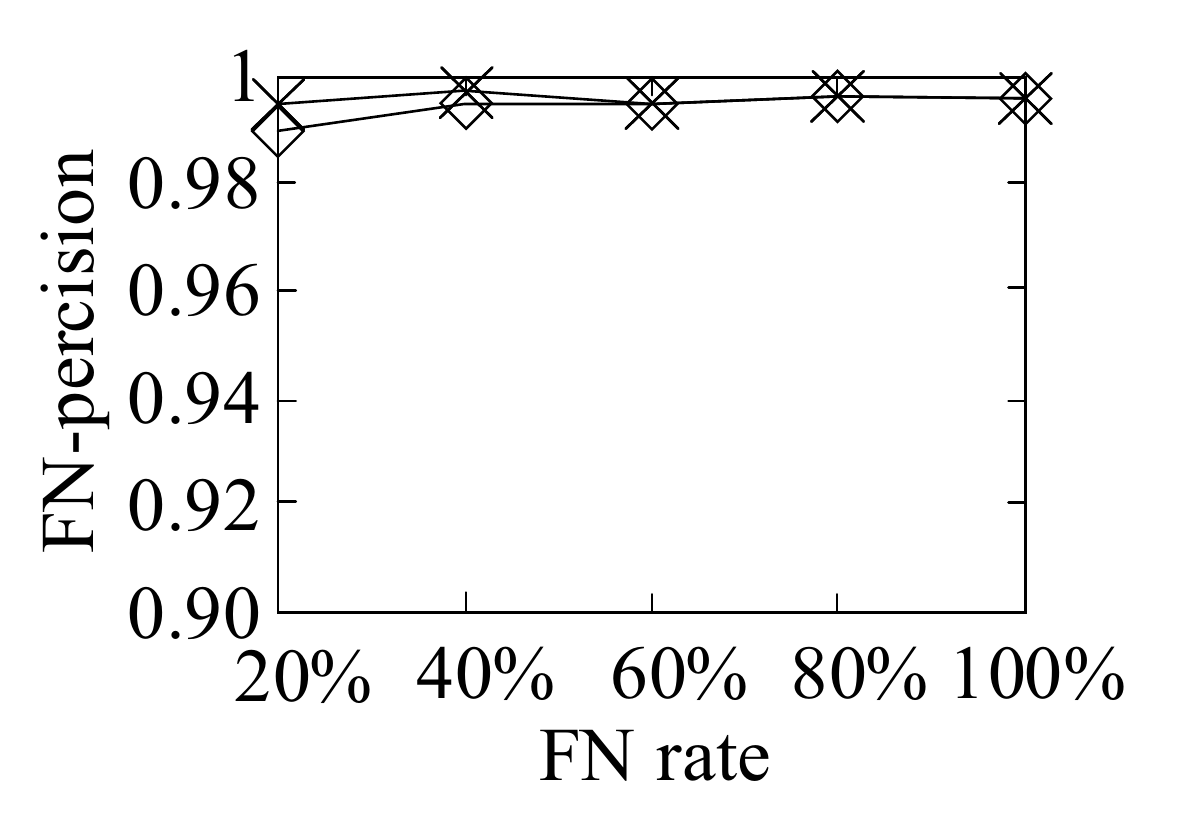}
  }\hspace*{-4mm}
  \subfigure[\emph{WN18}]{
   \includegraphics[width=1.8in]{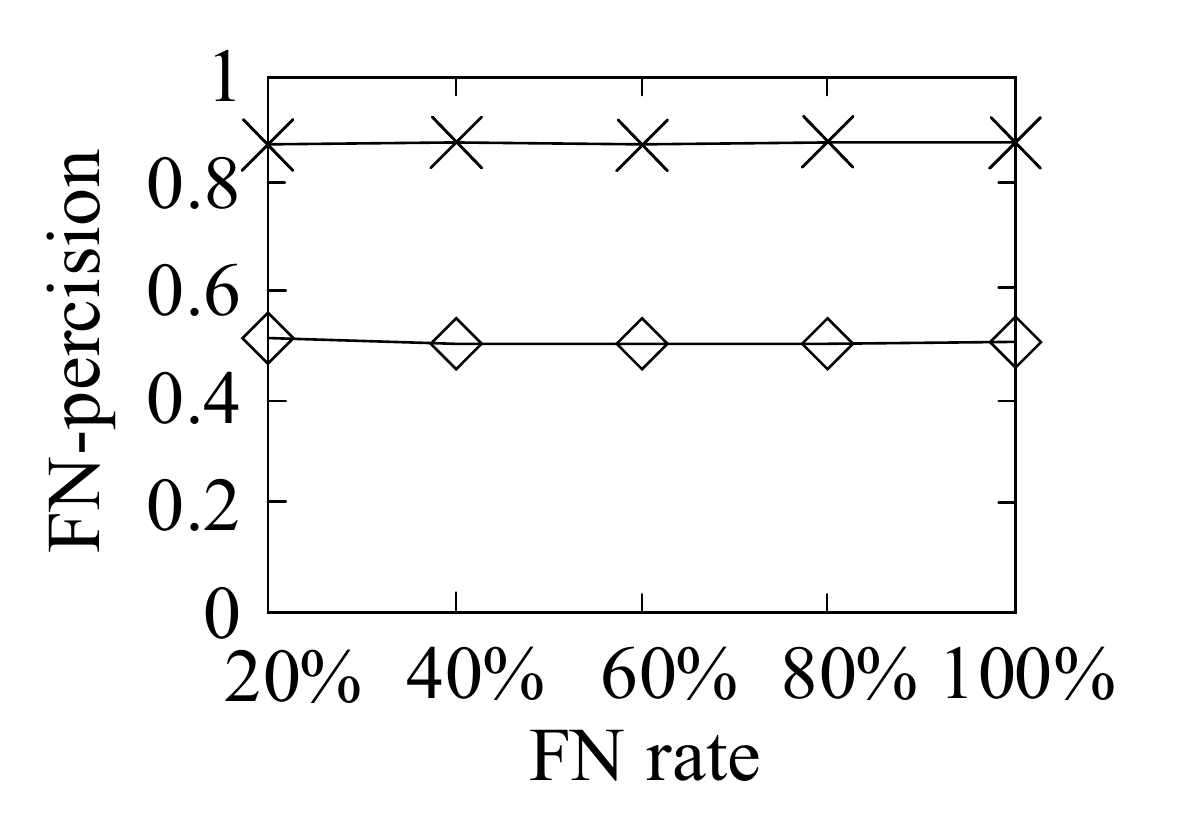}
  }\hspace*{-4mm}
\subfigure[\emph{WN18RR}]{
   \includegraphics[width=1.8in]{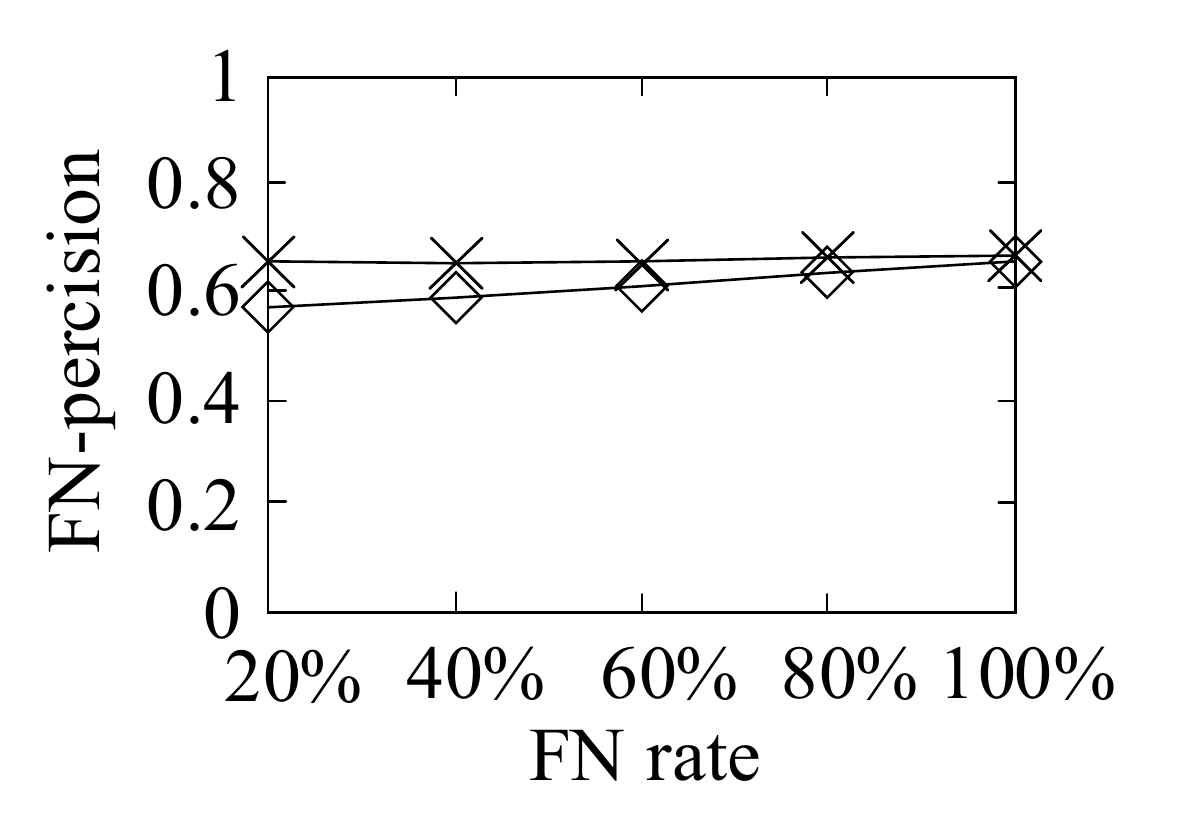}
  }
\vspace*{-4mm}
\caption{The performance of \textsf{PRO-repair} vs. false-negative rate}
\label{fig:fn-effect}
\vspace*{-2mm}
\end{figure*}

\begin{figure*}[t]
\centering
\includegraphics[width=0.38\textwidth]{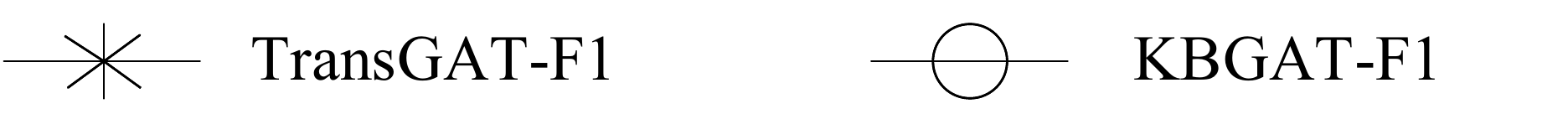}\\
\vspace*{-2mm}
\subfigure[\emph{UMLS}]{
   \includegraphics[width=1.75in]{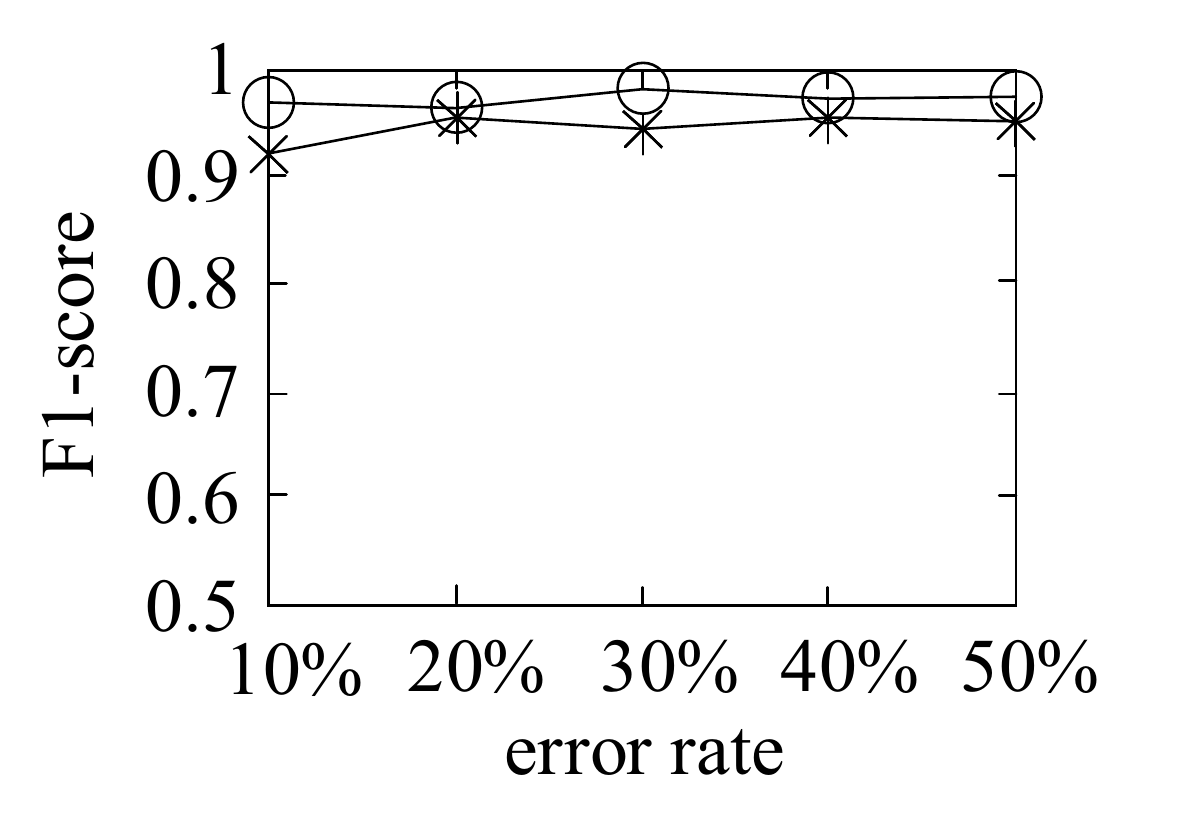}
  }\hspace*{-3mm}
\subfigure[\emph{Kinship}]{
   \includegraphics[width=1.75in]{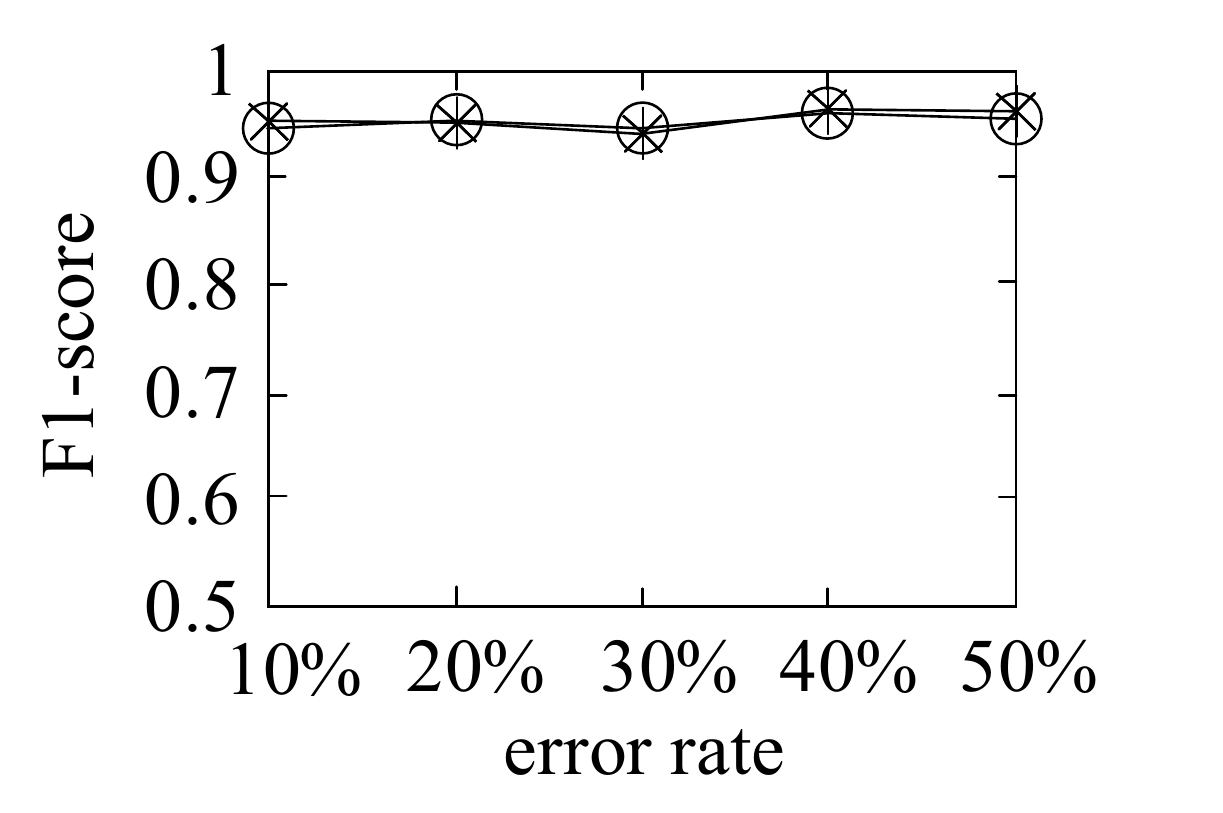}
  }\hspace*{-3mm}
  \subfigure[\emph{WN18}]{
   \includegraphics[width=1.75in]{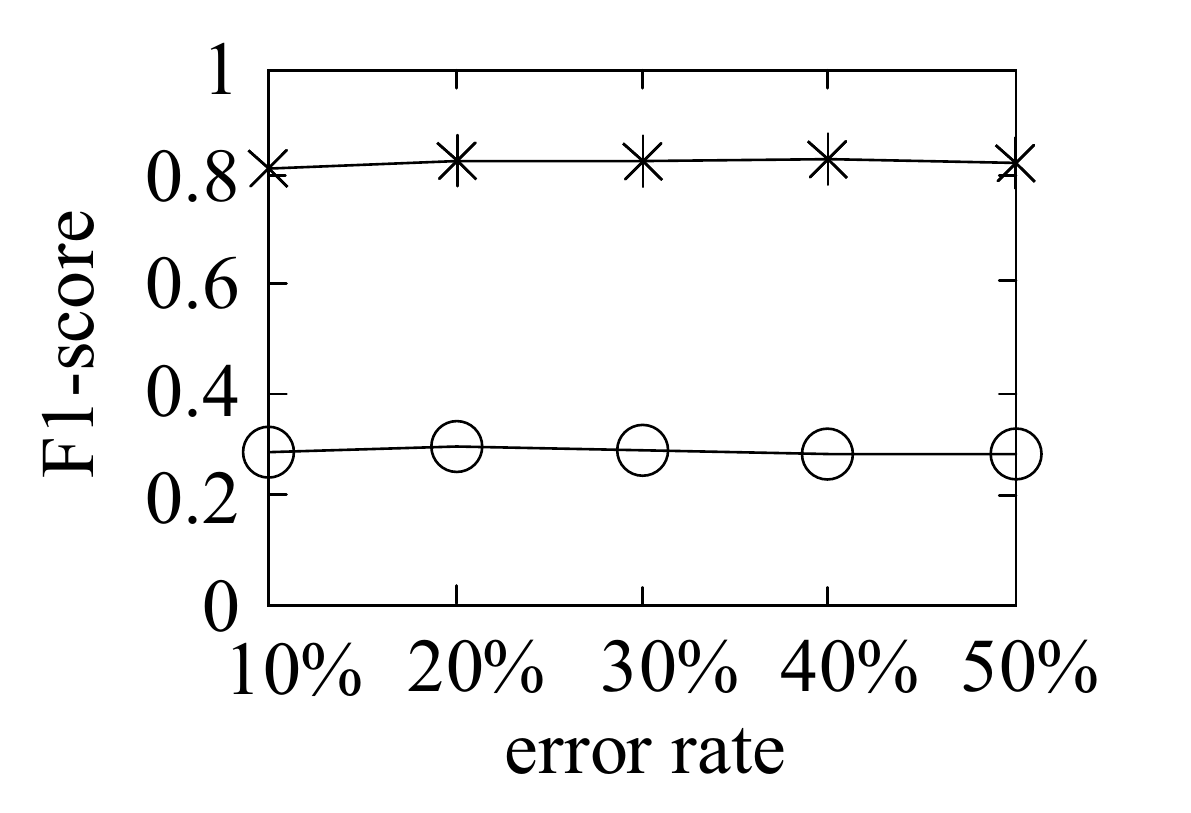}
  }\hspace*{-3mm}
\subfigure[\emph{WN18RR}]{
   \includegraphics[width=1.75in]{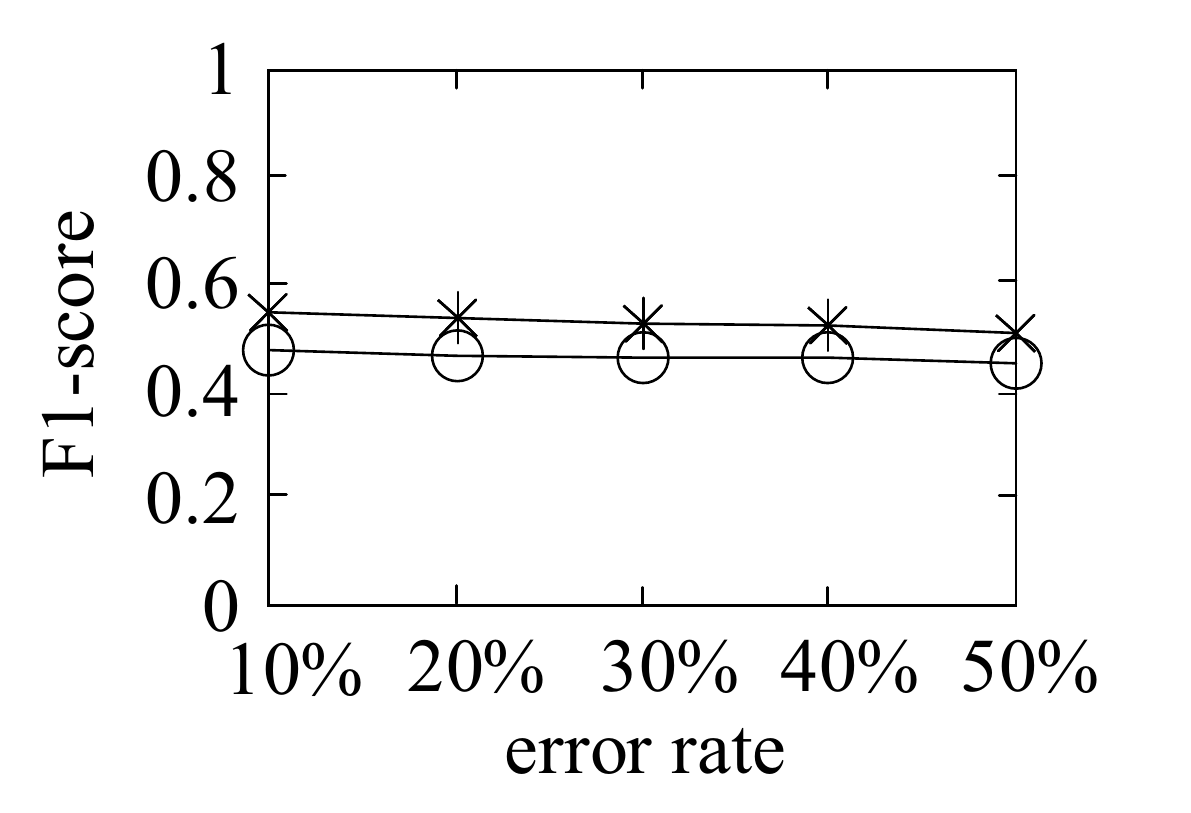}
  }
\vspace*{-4mm}
\caption{The performance of \textsf{KGClean} and its competitors under different error rates}
\label{fig:end-to-end-error-rate}
\vspace*{-3mm}
\end{figure*}

\noindent
\textbf{The Impact of Error Rate.}
We study the accuracy of \textsf{PRO-repair} by varying the error rate from 10\% to 50\%, and report the F1-score in Figure~\ref{fig:pro-repair-error-rate}. Here, we use F1-score, which is the harmonic mean between precision ($P$) and recall ($R$), i.e., $F1 = \frac{2 \cdot P \cdot R}{P+R}$. Let $TP$ refers to the number of repaired values of triplets that match the ground truth (note that, here $TP$ is different from TP defined in Section~\ref{subsec:al-detect}). Then, precision $P$ is defined as the fraction of $TP$ over the total number of the values being repaired; and recall $R$ is defined as the fraction of $TP$ over the total number of errors.  Note that, to eliminate the negative impact of incorrect classification results on \textsf{PRO-repair}, all experiments in this study are performed under the assumption that all triplets are correctly classified.
First, our repairing algorithm is observed to maintain a stable performance when the error rate grows. Its resilience to the error rate is mainly contributed by the well-trained embedding model (i.e., \textsf{TransGAT}), which provides the ability to find the correct values via the propagation power of triplets.
Second, the cleaning result of \textsf{PRO-repair} on the WN18RR dataset is not as good as that of other datasets, including UMLS, Kinship, and WN18. This is because the KG of WN18RR is very sparse which has a significant number of nodes with no incoming edges, i.e., nodes with zero in-degree in KG~\cite{NathaniCSK19}. Consequently, it is hard to use the interactions between the nodes and their corresponding relationships or to gather their neighbors' information for learning accurate embeddings. As a result, the candidates' propagation power of each noisy triplet may become less reliable, which directly affects the quality of the repairing results. In addition, as expected, the \textsf{TransGAT}-based \textsf{PRO-repair} has better performance than the KBGAT-based method, consistently across all datasets. This is because \textsf{TransGAT} provides a more accurate score function, which is used in the propagation power, than that of KBGAT.


\noindent
\textbf{The Impact of Error Detecting.}
We also explore the impact of error detecting on the quality of \textsf{PRO-repair}'s output.
Since \textsf{KGClean} is restricted to execute \textsf{PRO-repair} only for triplets that are identified as potentially erroneous by the \textsf{AL-detect} strategy, it is natural that \textsf{PRO-repair} can not repair undetected noisy triplets.
We study the impact of \emph{false-negative triplets}, which are classified as errors but are actually not.
We define a new variable \emph{false-negative rate} as the proportion of false-negative triplets to the total clean triplets, and vary the false-negative rate from 20\% to 100\%. A new metric \emph{FN-precision} is introduced for evaluation. The FN-precision is denoted as the fraction of the truly repaired false-negative triplets over the total number of the false-negative triplets.
We report our results in Figure~\ref{fig:fn-effect}. We see that \textsf{PRO-repair} is not sensitive to the false-negative rate.
This is because the reliable embeddings learned by \textsf{TransGAT} ensure that the propagation power of a clean triplet is greater than that of its associated repairing candidates. Therefore, it is unlikely to repair the false-negative triplets with other erroneous candidates incorrectly.

\subsection{End-to-End Performance}

Last but not the least, we evaluate the performance of our \textsf{KGClean} based on our proposed \textsf{TransGAT} and that based on KBGAT for cleaning errors in KGs, respectively, by varying the error rate from 10\% to 50\%.
Note that, we do not evaluate the performance of \textsf{KGClean} using the TransE model. This is because the structure of TransE is different from that of \textsf{TransGAT}, e.g., the embeddings of TransE do not contain information of multi-hop neighbors, and thus, the \textsf{PRO-repair} strategy is not applicable to the TransE-based cleaning framework. In addition, as reported in  Section~\ref{sec:analyse_transGAT}, the embedding results of TransE are far more inferior to \textsf{TransGAT}'s embeddings. Again, we adopt F1-score as the performance metric.

The overall results are plotted in Figure~\ref{fig:end-to-end-error-rate}. From the results, we can observe that the performance of both versions of \textsf{KGClean} is not sensitive to the error rate.
The reason is that we adopt KG embeddings as the external information for cleaning. In particular, both error detecting and error repairing phases of \textsf{KGClean} rely on the score functions of the graph embedding models, which are only related to the training set but not the test set. Therefore, no matter how we change the error rate of the test set, the performance remains almost unchanged.
As expected, \textsf{KGClean} framework based on \textsf{TransGAT} outperforms the one based on KBGAT.
This is because the embeddings of \textsf{TransGAT} can preserve the semantic information of triplets more accurately than the embeddings of KBGAT, as reported in Section~\ref{sec:analyse_transGAT}. Under different error rates, the slight oscillation of the F1-score is caused by the randomness of error generation. One exception is the UMLS dataset, where KBGAT-based \textsf{KGClean} achieves higher F1 score than the \textsf{TransGAT}-based \textsf{KGClean}. The reason attributes to the unreliable prediction of \textsf{AL-detect} when using KBGAT embeddings, as explained in Section~\ref{subsec:al-detect}.

\section{Conclusions}
\label{sec:conclusion}
We propose a novel embedding powered knowledge graph cleaning framework \textsf{KGClean} in this paper.
It first learns data representations by our presented \textsf{TransGAT}, an effective knowledge graph embedding model, which gathers the neighborhood information of each entity and incorporates the interactions between entities and relationships for casting data to vector spaces with rich semantics. Then, \textsf{TransGAT} uses an active-learning-based classification model to identify noisy triplets from a dirty knowledge graph. Next, \textsf{TransGAT} fixes erroneous values within the set of noisy triplets according to a novel concept of propagation power. Extensive experimental results on four typical knowledge graphs demonstrate the effectiveness of \textsf{KGClean}.

\balance

\bibliographystyle{IEEEtran}
\bibliography{KGClean}


\end{document}